# Computational Quantum Secret Sharing

Alper Çakan
Carnegie Mellon University
acakan@andrew.cmu.edu

Vipul Goyal
NTT Research &
Carnegie Mellon University
vipul@cmu.edu

Chen-Da Liu-Zhang[*]
NTT Research
chen-da.liuzhang@ntt-research.com

João Ribeiro[*]
NOVA LINCS
NOVA School of Science and Technology
joao.ribeiro@fct.unl.pt

**Abstract**

Quantum secret sharing (QSS) allows a dealer to distribute a secret *quantum state* among a set of parties in such a way that certain authorized subsets can reconstruct the secret, while unauthorized subsets obtain no information about it. Previous works on QSS for general access structures focused solely on the *existence* of perfectly secure schemes, and the share size of the known schemes is necessarily exponential even in cases where the access structure is computed by polynomial size monotone circuits. This stands in stark contrast to the classical setting, where *polynomial-time* computationally-secure secret sharing schemes have been long known for all access structures computed by polynomial-size monotone circuits under standard hardness assumptions, and one can even obtain shares which are much shorter than the secret (which is impossible with perfect security).

While QSS was introduced over twenty years ago, previous works only considered information-theoretic privacy. In this work, we initiate the study of *computationally-secure* QSS and show that computational assumptions help significantly in building QSS schemes, just as in the classical case. We present a simple compiler and use it to obtain a large variety results: We construct *polynomial-time* computationally-secure QSS schemes under standard hardness assumptions for a rich class of access structures. This includes many access structures for which previous results in QSS necessarily required exponential share size. In fact, we can go even further: We construct QSS schemes for which the size of the quantum shares is significantly smaller than the size of the secret. As in the classical setting, this is impossible with perfect security.

We also apply our compiler to obtain results beyond computational QSS. In the information-theoretic setting, we improve the share size of perfect QSS schemes for a large class of $n$-party access structures to $1.5^{n+o(n)}$, improving upon best known schemes and matching the best known result for general access structures in the classical setting. Finally, among other things, we study the class of access structures which can be efficiently implemented when the quantum secret sharing scheme has access to a given number of copies of the secret, including all such functions in P and NP.

---

[*]Part of the work was done while at Carnegie Mellon University.

# Contents

# 1 Introduction

Classical secret sharing [Bla79, Sha79] is a fundamental cryptographic primitive which allows one to share a classical bit-string (the secret) among $n$ parties so that (i) only authorized subsets of parties can reconstruct the secret, and (ii) unauthorized subsets have essentially no information about the secret. The associated class of authorized sets, called the *access structure*, is defined by a monotone function $f : \{0,1\}^n \to \{0,1\}$ with a set $P \subseteq [n]$ being authorized if and only if $f(v^P) = 1$, where $v^P \in \{0,1\}^n$ is the characteristic vector of $P$ satisfying $v_i^P = 1$ exactly when $i \in P$. For the sake of convenience, we will often write $f(P)$ for $f(v^P)$ when the context is clear.

Secret sharing has found several applications in cryptography, see the extensive survey of Beimel [Bei11] for a discussion of such applications. Motivated by these applications, it is important to design secret sharing schemes realizing a given monotone function which are as efficient as possible, be it in terms of requiring polynomial-time sharing and reconstruction procedures or, more modestly, requiring that the resulting shares be as short as possible. Blakley [Bla79] and Shamir [Sha79] originally described efficient secret sharing schemes for *threshold* monotone functions (where a set $P$ is authorized if and only if $|P| \geq t$ for a given threshold $t$). A long line of research over the past 40 years has significantly extended and complemented these results.

Of particular importance to us, it is known how to exploit different computational hardness assumptions to obtain efficient *computational* secret sharing schemes realizing a broad class of monotone functions, where by *efficient* we mean that the scheme enjoys polynomial-time sharing and reconstruction. In particular, such schemes also have polynomial sized shares. Early work of Yao [Yao89, VNS$^+$03] described families of efficient *computational* secret sharing schemes realizing all functions in monotone P, i.e., all sequences of monotone functions $(f_n)_{n \in \mathbb{Z}^+}$ with $f_n : \{0,1\}^n \to \{0,1\}$ that are computed by a sequence of poly($n$)-size monotone circuits,[*] based on the existence of one-way functions. A more recent work by Komargodski, Naor, and Yogev [KNY14] succeeded in designing such efficient computational schemes realizing all functions in[†] mNP assuming witness encryption for NP and one-way functions.[‡] Following this, Bartusek and Malavolta [BM21] designed efficient computational schemes for *classical* secrets realizing all monotone functions in QMA assuming witness encryption for this class.

Given the advent of quantum computing, it is natural to consider the construction of schemes for sharing a *quantum state*, as opposed to a classical bitstring. This problem was first considered in [HBB99, KKI99, CGL99] for some specific monotone functions. Follow-up works by Gottesman [Got00] and Smith [Smi00] showed how to design quantum secret sharing schemes for all allowable monotone functions.[§] Of particular relevance, Smith [Smi00] constructed quantum secret sharing schemes realizing monotone functions $f$ whose total share size is equal to the size of the smallest *monotone span program* computing $f$, thus generalizing a seminal classical result of Karchmer and Wigderson [KW93] to the quantum setting.

Remarkably, the result of Smith [Smi00] still remains the state-of-the-art for quantum secret sharing realizing broad classes of monotone functions more than 20 years later, and it leaves significant loose ends. In fact, it is now known that there are functions in monotone P which only have exponentially large associated monotone span programs [RPRC16, PR17]. This means that with the current methods the best quantum secret sharing scheme realizing such monotone func-tions requires exponential share size. In contrast, as mentioned before, we have efficient computational *classical* secret sharing schemes realizing all such functions under standard hardness assumptions [Yao89, VNS⁺03]. Moreover, if we are willing to upgrade our hardness assumptions, then we know such efficient classical schemes for the much broader class of monotone functions in mNP [KNY14].

---

[*]Throughout this paper, we make the parametrization of the sequence of monotone functions in terms of $n$ implicit when clear from context.

[†]We denote by mNP the class of monotone functions in NP.

[‡]The notion of secret sharing for functions in mNP requires that the reconstructor receive not only an authorized subset of shares, but also a polynomial-size witness certifying that this subset is authorized.

[§]Not all monotone functions allow a quantum secret sharing scheme. We discuss this in more detail later.

## 1.1 A summary of our contributions

Given the state of affairs above, the following questions arise naturally:

> *Can we use computational hardness assumptions to significantly expand the class of monotone functions that can be realized by* efficient *quantum secret sharing schemes? Furthermore, can we improve the share size of schemes for an even broader class of monotone functions?*

We make significant progress in this direction via a new and streamlined approach. While the concept of quantum secret sharing has been around for over twenty years, the work so far has only considered the notion of information-theoretic privacy for such schemes, in contrast to the classical setting. In this work, we initiate the study of *computationally-secure quantum secret sharing*. By leveraging standard hardness assumptions, we show how a conceptually simple compiler utilizing the idea of hybrid encoding allows us to obtain schemes which are far more efficient than those constructed using the current methods. Our simple compiler allows us to obtain a quantum secret sharing scheme from a classical secret sharing scheme and a quantum erasure correcting code. Using this compiler, we design quantum secret sharing schemes with various desirable properties by lifting well-known classical results to the quantum setting. In particular, we are able to lift the results of Yao [Yao89, VNS⁺03] and Komargodski-Naor-Yogev [KNY14] to the quantum setting for a broad class of monotone functions which we show inherits many relevant properties from general monotone functions. Moreover, we are also able to use our general approach to lift many other results from classical secret sharing to the quantum setting, such as computational secret sharing of long messages with short shares and general perfect secret sharing with share size breaking the circuit barrier. Finally, we are able to obtain efficient schemes for any function in monotone P, given sufficiently many (at most $n$) copies of the secret.

**The key difficulty.** One major difficulty that separates quantum from classical secret sharing is the *no-cloning theorem* [WZ82], which precludes copying unknown quantum states. This means that basic techniques, such as giving copies of the same component to several parties, cannot be exploited. In other words, there are no quantum secret sharing schemes realizing the OR function. Consequently, the approaches behind many fundamental classical results cannot be directly extended to the quantum setting, and so this lifting requires different ideas. We proceed to describe our contributions and formal results in more detail. Our results follow from a simple generic compiler, using the general paradigm of hybrid encoding, from classical to quantum secret sharing that we design and analyze.

**Heavy monotone functions.** Due to the no-cloning theorem, quantum secret sharing schemes are only able to realize what we call *no-cloning* monotone functions [CGL99]. These are monotone functions $f$ with the property that $f(P) = 1$ implies $f(\overline{P}) = 0$, i.e., the complement of an authorized set is unauthorized. As discussed above, we know how to construct quantum secret sharing schemes realizing all no-cloning monotone functions [Got00, Smi00], but the state-of-the-art share size for all such functions $f$ corresponds to the size of the smallest monotone span program computing $f$, which may be extremely large even for "simple" no-cloning monotone functions $f$ in monotone P.

We first focus on a natural subclass of no-cloning monotone functions which we call *heavy* monotone functions.

**Definition 1** (Heavy function). *A monotone function $f : \{0,1\}^n \to \{0,1\}$ is said to be $t$-heavy if for any set $P \subseteq [n]$ with $f(P) = 1$, we have $|P| \geq t$. When $t \geq \lfloor \frac{n}{2} \rfloor + 1$, we simply say that $f$ is* heavy.

Equivalently, $t$-heavy monotone functions correspond exactly to the class of monotone functions with *minimal authorized sets*¶ of size at least $t$. Note also that a $t$-out-of-$n$ threshold function is a special case of a $t$-heavy function.

A $t$-heavy function with $t > n/2$ satisfies the no-cloning property, and thus can be realized by a quantum secret sharing scheme. Naturally, one may wonder whether the class of heavy monotone functions is interesting. For example, it could be the case at first sight that all heavy monotone functions are computed by polynomial-size monotone span programs, in which case we would already know efficient quantum secret sharing schemes for all such functions via Smith's construction [Smi00]. We show that this is, in fact, not the case: Heavy monotone functions inherit the complexity of *arbitrary* monotone functions, as made precise in the following result (see Section 9 for the proof).

**Proposition 1.** *Let $\mathsf{mSP}(f)$ and $\mathsf{mC}(f)$ denote the size of the smallest monotone span program and monotone circuit computing $f$, respectively. Then, for every monotone function $f : \{0,1\}^n \to \{0,1\}$ there exists a heavy monotone function $f' : \{0,1\}^{2n} \to \{0,1\}$ such that $\mathsf{mSP}(f') \geq \frac{\mathsf{mSP}(f)}{2n}$ and $\mathsf{mC}(f') \leq \mathsf{mC}(f) + n$. Moreover, if $f$ is in $\mathsf{mNP}$, then so is $f'$.*

In words, Proposition 1 states that for every monotone function $f$ there is a corresponding heavy monotone function $f'$ with essentially the same complexity in terms of monotone span programs and monotone circuits. Combining Proposition 1 with recent results from [RPRC16, PR17] leads to the following corollary.

**Corollary 1.** *There exist heavy monotone functions in* monotone P *which require monotone span programs of size $\exp\left(n^{\Omega(1)}\right)$.*

This corollary is very relevant in the context of secret sharing. It shows that there are *heavy* monotone functions for which we have efficient computational *classical* secret sharing schemes via Yao's construction, but for which the best known method [Smi00] for constructing *quantum* secret sharing schemes requires exponential share size.

**Computational quantum secret sharing of long messages with short shares.** As our first contribution, we consider the task of sharing secrets consisting of multiple qubits at once. Naively, this can be accomplished by sharing each qubit in parallel. However, we would like to do considerably better.

In the classical setting, Krawczyk [Kra94] showed that we can share a sufficiently long secret bitstring *using shares that are much shorter than the secret* under standard hardness assumptions via a basic technique. Moreover, it is easy to see that this is impossible to achieve in the case of perfect secret sharing [KGH83]. Using our general approach, we obtain a quantum analogue of Krawczyk's result, thus showing that we can share a high-dimensional secret quantum state using quantum shares of much lower dimension than the secret under standard post-quantum hardness assumptions. As in the classical setting, this is impossible to achieve with perfect security. Gottesman [Got00] showed that, in this case, shares of important parties (i.e., parties whose removal from some authorized set make it unauthorized) must be at least as large as the secret. Therefore, in particular, the above cannot be achieved by perfectly secure quantum secret sharing schemes even for the simplest case of threshold monotone functions. Furthermore, the currently known quantum secret sharing schemes can only achieve exponential share size per message size ratio for some functions whereas below we show that we can achieve a share size per message size ratio below 1 for sufficiently long messages.

---
¶We say that $P$ is a *minimal authorized set for $f$* if $f(P) = 1$ but $f(S) = 0$ for all strict subsets $S \subsetneq P$.

We now provide some more details. Suppose we wish to share a secret composed by $m$ qubits. Then, our quantity of interest is the *information ratio* of a quantum secret sharing scheme [Bei11], which is given by

$$\frac{\max_{i \in [n]} |S_i|}{m},$$

where $|S_i|$ denotes the number of qubits used to describe the $i$-th share of the scheme. Then, our goal is to design efficient quantum secret sharing schemes whose information ratio is as small as possible as a function of the secret size $m$ and the number of parties $n$. As mentioned above, information-theoretic schemes always have information ratio at least 1 [Got00], and so we must use computational assumptions to break this barrier. The following theorem is a notable special case of our general approach.

**Theorem 1.** *If $f$ is a $t$-heavy monotone function, with $t > n/2$, computed by monotone circuits of size $O(n^d)$, then there is an efficient computational quantum secret sharing scheme realizing $f$ with asymptotic information ratio at most $\frac{32}{2t-n}$ for secrets composed of at least $m = \Omega(n^{cd})$ qubits for a universal constant $c > 0$ based on the existence of post-quantum secure one-way functions.*

In particular, observe that when $t > (\frac{1}{2} + \delta)n$ for an arbitrary constant $\delta > 0$, Theorem 1 guarantees that the information ratio is not only well below 1 but actually behaves as $O(1/n)$ when $n$ is large enough. This is optimal up to the constant factor, since the sum of the sizes of all shares must be at least $m \geq 1$.

**Efficient computational quantum secret sharing.** As our second result, we show that we can leverage computational hardness assumptions to obtain significantly more efficient quantum secret sharing schemes for heavy monotone functions, even when sharing a single qubit.

**Theorem 2.** *If $f$ is a heavy monotone function in monotone P, then there is an efficient computational quantum secret sharing scheme realizing $f$ based on the existence of post-quantum secure one-way functions.*

**Theorem 3.** *If $f$ is a heavy monotone function in mNP, then there is an efficient computational quantum secret sharing scheme realizing $f$ based on the existence of post-quantum secure witness encryption for NP and one-way functions.*

Note that, by Corollary 1, it follows that both Theorems 2 and 3 provide an *exponential* improvement on the efficiency of known quantum secret sharing schemes for heavy monotone functions.

Interestingly, computational quantum secret sharing had not been studied until our work. Furthermore, there is a curious phenomenon with respect to computational privacy in the quantum setting: Because of the no-cloning theorem, there are monotone functions for which correctness of the quantum secret sharing scheme gives perfect privacy for free, and so computational assumptions cannot be used to obtain improved schemes realizing such monotone functions. More precisely, if $f$ is *self-dual*, meaning that $f(P) = 1$ if and only if $f(\overline{P}) = 0$, then correcting erasures on $\overline{P}$ implies (via the no-cloning theorem) that the shares corresponding to the subset of parties $\overline{P}$ yield no information about the secret qubit [CGL99]. Therefore, perfect reconstruction implies perfect privacy in this case. On the other hand, our results above show that computational assumptions *can* be helpful in obtaining efficient quantum secret sharing schemes realizing a broader class of monotone functions.

We remark that our results extend beyond the class of heavy monotone functions. We discuss these extensions in detail later in this section.

**Beyond computational secret sharing: Perfect quantum secret sharing with share size breaking the circuit size barrier.** We also extend our approach to obtain new results beyond computational quantum secret sharing. Until recently, the state-of-the-art classical *perfect* secret sharing schemes for arbitrary $n$-party monotone functions required share size $\Omega(2^n/\sqrt{n})$ [BL88] – the so-called *circuit-size barrier*. However, a recent groundbreaking line of research [LVW18, LV18, ABF+19, ABNP20, AN21] has succeeded in constructing classical perfect secret sharing schemes for arbitrary monotone functions with share size $1.5^{n+o(n)}$, well below the circuit size barrier. In this work, we obtain a quantum analogue of this result.

**Theorem 4.** *If $f : \{0,1\}^n \to \{0,1\}$ is a heavy monotone function, then there is a* perfect *quantum secret sharing scheme realizing $f$ with a total share size of $1.5^{n+o(n)}$ classical bits and $O(n \log n)$ qubits.*

Observe that Theorem 4 is a significant improvement over the previous known results for information-theoretic quantum secret sharing [Got00, Smi00]. In fact, if all heavy monotone functions were computed by monotone span programs of size less than $1.5^{n+o(n)}$, then Proposition 1 implies that the same would hold for *all monotone functions* – a major improvement over currently known results. Therefore, the previous constructions from [Got00, Smi00] require much larger total share size for some heavy monotone functions than Theorem 4.

**Bypassing the no-cloning theorem: Quantum secret sharing with multiple copies.** As we have seen above, not all monotone functions can be realized by standard quantum secret sharing schemes due to the no-cloning theorem. Therefore, it is natural to wonder what kind of additional assumptions are necessary to bypass this barrier and design quantum secret sharing schemes for a wider range of monotone functions. Arguably, one of the most reasonable directions is to assume we have access to several copies of the quantum state to be shared. For example, this makes sense in the setting of multiparty computation, where each party has a classical description of their quantum input and thus may create as many copies as it wants. The following question arises naturally from this discussion:

> *How many copies are required to design quantum secret sharing schemes realizing* all *monotone functions in* monotone P*?*

Chien [Chi20] considered this question for the special case of threshold monotone functions. More precisely, he gave a scheme showing (without proof of security) that

$$\max(1, n - 2t + 2)$$

copies of the quantum secret are sufficient to obtain a $t$-out-of-$n$ quantum secret sharing scheme. Yet, the efficiency of such schemes was not considered. We exploit our approach to show that the $t$-out-of-$n$ threshold function is the most demanding among all $t$-heavy monotone functions, and then construct efficient quantum secret sharing schemes for threshold functions given multiple copies, which allows us to settle (a more general version of) the question above in both the computational and information-theoretic settings.

**Theorem 5.** *A total of $\max(1, n - 2t + 2)$ copies of the quantum secret are sufficient to obtain efficient computational quantum secret sharing schemes realizing all $t$-heavy monotone functions $f$ in* monotone P *assuming the existence of post-quantum secure one-way functions.*

Observe that every monotone function $f$ is $t$-heavy when $t$ is the minimum size of its authorized sets. Therefore, given sufficiently many copies, we are able to obtain efficient computational quantum secret sharing schemes for any function $f \in$ monotone P.

**Corollary 2.** *For any monotone function $f : \{0,1\}^n \to \{0,1\}$ in* monotone P *there is an efficient computational quantum secret sharing scheme using at most $n$ copies of the secret realizing $f$ based on the existence of post-quantum secure one-way functions.*

If we are willing to settle for larger share size, we can obtain an analogous result for all $t$-heavy functions, including those that are not in monotone P.

**Theorem 6.** *A total of $\max(1, n - 2t + 2)$ copies of the quantum secret are sufficient to obtain perfect quantum secret sharing schemes realizing all $t$-heavy monotone functions $f$ over $n$ parties.*

**Beyond heavy monotone functions.** Our techniques can be applied to classes of monotone functions greatly generalizing the class of heavy monotone functions. Naturally, such classes of functions also inherit the hardness properties of arbitrary monotone functions detailed in Proposition 1. We give two interesting examples.

*Weighted-heavy monotone functions.* As a natural generalization of heavy functions, we consider *weighted-heavy* functions. Intuitively, in a weighted-heavy monotone function each party is assigned a weight – the function must evaluate to 0 for a set of parties if the sum of their weights is below some threshold $t$, and is otherwise unconstrained. Such functions can also be seen as natural extensions of weighted threshold functions.

**Definition 2** (Weighted heavy function). *Let $w : [n] \to \mathbb{N}$ be an integer weight function. A monotone function $f : \{0,1\}^n \to \{0,1\}$ is said to be $(w,t)$-weighted-heavy if for any set $P \subseteq [n]$ with $f(P) = 1$ we have $\sum_{i \in P} w(i) \geq t$. Moreover, we call $W = \sum_{i=1}^n w(i)$ the* total weight *of $f$. If $f$ is $(w,t)$-weighted-heavy for some $w$ and $t \geq \lfloor \frac{W}{2} \rfloor + 1$, we simply call it $w$-weighted-heavy or just* weighted-heavy *if $w$ is clear from context.*

Note that all $(w,t)$-weighted heavy monotone functions with $t > W/2$ satisfy the no-cloning property, and we may see $t$-heavy functions as $(w,t)$-weighted heavy functions with $w(i) = 1$ for all $i \in [n]$. One of the reasons why this is interesting is the following proposition, which is related to Proposition 1 connecting general and heavy monotone functions and shows that weighted heavy functions strictly generalize heavy and weighted threshold functions (see Section 10.1 for a proof).

**Proposition 2.** *There are families of $w$-weighted heavy monotone functions with total weight $W = \mathrm{poly}(n)$ which are neither heavy nor weighted threshold functions with $\mathrm{poly}(n)$ weights. Moreover, there exist such functions which are also in* monotone P *but require monotone span programs of size $\exp\left(n^{\Omega(1)}\right)$.*

In general, we show that if the sum of the weights $W$ is polynomial in the number of parties $n$ and $t > W/2$, as is the case for the family of functions in Proposition 2, then we can construct efficient computational quantum secret sharing schemes for weighted-heavy monotone functions in monotone P, generalizing Theorem 2.

**Theorem 7.** *If $f : \{0,1\}^n \to \{0,1\}$ is a $(w,t)$-weighted-heavy monotone function in monotone P with total weight $W = \text{poly}(n)$ and threshold $t > W/2$, then there is an efficient computational quantum secret sharing scheme realizing $f$ based on the existence of post-quantum secure one-way functions.*

*Trees of weighted-heavy monotone functions.* Our techniques can be further applied to constant depth trees that are composed of gates computing weighted heavy functions. Concretely, consider the family $\mathcal{F}$ of weighted heavy functions in monotone P with $\text{poly}(n)$ total weight that satisfy the no-cloning property. Then, we design efficient computational quantum secret sharing schemes realizing any monotone function computed by constant depth trees consisting of gates in $\mathcal{F}$ with fan-in at most $n$.

**Theorem 8.** *Let $\mathcal{F}$ be the family of weighted heavy functions in monotone P with $\text{poly}(n)$ total weight and fan-in at most $n$. If $f : \{0,1\}^n \to \{0,1\}$ is computed by a constant depth polynomial size tree composed of gates in $\mathcal{F}$, then there is an efficient computational quantum secret sharing scheme realizing $f$ based on the existence of post-quantum secure one-way functions.*

Making a parallel with Proposition 2, we are able to show that trees of weighted-heavy monotone functions strictly generalize weighted-heavy monotone functions – in fact, depth 2 is already sufficient for this (see Section 10.2 for a proof). Recall that we had already seen in Proposition 2 that the latter strictly generalize heavy monotone functions and weighted threshold functions.

**Proposition 3.** *There are families of functions $f : \{0,1\}^n \to \{0,1\}$ which satisfy all of the following at once:*

- *$f$ is not weighted-heavy;*

- *$f \in$ monotone P;*

- *$f$ requires monotone span programs of size $\exp\left(n^{\Omega(1)}\right)$;*

- *$f$ can be represented as a depth-$2$ polynomial size tree of weighted-heavy monotone functions where each gate is in monotone P and has polynomially bounded total weight.*

### 1.2 Technical overview

We now discuss our techniques and approach in more detail.

**The general compiler.** The starting point in our approach is a simple and versatile compiler which exploits the fact that we can perfectly encrypt a quantum state using a classical key. This compiler shares some ideas with Krawczyk's classical scheme [Kra94], and similar techniques have been exploited in an orthogonal direction to reduce the number of quantum shares in perfect quantum secret sharing schemes realizing threshold monotone functions [NMI01, FG12]. It combines a classical secret sharing scheme with the quantum one-time pad (QOTP) [AMTD00] and a quantum erasure-correcting code.

On a high level, our compiler works by perfectly encrypting a state using QOTP. Then, since the keys are classical, we establish the security according to the desired access structure by simply secret sharing the keys using an efficient classical scheme. Then, to allow any authorized set of parties to reconstruct the secret, we *distribute* the encrypted state using a quantum erasure correcting code. We note that the "hybrid encoding" approach we undertake here is prevalent in quantum computing. Other examples of this approach have appeared in the literature (see Section 1.3 for details).

Quantum erasure-correcting codes (QECCs) are a quantum analogue of classical erasure-correcting codes. Intuitively, a length-$n$ QECC of dimension $k$ maps an input quantum state $\rho$ over $k$ qubits into a higher-dimensional quantum state $E_\rho$ over $n$ qubits with the property that even if some qubits at known positions of $E_\rho$ are subjected to any error, then it is still possible to perfectly recover $\rho$ from the corrupted quantum codeword. General constructions with good parameters have been known since at least the seminal work of Calderbank and Shor [CS96] and Steane [Ste96].

More precisely, for a given monotone function $f : \{0,1\}^n \to \{0,1\}$, suppose we have access to a *classical* secret sharing scheme $\mathsf{SS} = (\mathsf{SS.Share}, (\mathsf{SS.Rec}_P)_{P \subseteq [n]})$ realizing $f$ and a QECC $\mathsf{QC} = (\mathsf{QC.Enc}, \mathsf{QC.Rec})$ with $n$ *components* "realizing" some monotone function $f'$ satisfying $f'(x) \geq f(x)$ for all $x \in \{0,1\}^n$. That is, $\mathsf{QC}$ corrects erasures in the complement of all sets $P$ such that $f'(P) = 1$. Then, our general compiler proceeds as follows on input an arbitrary qubit $\rho$:

1. Sample a classical key $k = (k_1, k_2) \leftarrow \{0,1\}^2$;

2. Encrypt $\rho$ with QOTP using key $k$, yielding the perfectly encrypted qubit

$$E_\rho = \mathsf{OTPEnc}(\rho, k) = X^{k_1} Z^{k_2} \rho (X^\dagger)^{k_1} (Z^\dagger)^{k_2},$$

   where $X$ and $Z$ are Pauli gates;

3. Share $k$ using $\mathsf{SS.Share}$, yielding classical shares $(S_1, \ldots, S_n)$;

4. Encode $E_\rho$ using $\mathsf{QC.Enc}$, yielding entangled quantum systems $(E_1, \ldots, E_n)$;

5. Set $(S_i, E_i)$ as the final share of the $i$-th party.

Note that this compiler can be easily generalized to states of arbitrary dimension. Naturally, we need to assume $f'$ (and hence $f$) satisfies the *no-cloning* property ($f(P) = 1 \implies f(\overline{P}) = 0$). This must be so that we can employ an appropriate QECC. Nevertheless, there is no loss of generality, since, as we have discussed before, quantum secret sharing is impossible when $f$ does not satisfy this property, due to the no-cloning theorem [CGL99].

We claim that the resulting scheme is a quantum secret sharing scheme realizing $f$. Let $(S_P, E_{\rho,P})$ denote the set of shares of $\rho$ that belong to a subset of parties $P$. It is straightforward to show that we can reconstruct $\rho$ from $(S_P, E_{\rho,P})$ when $f(P) = 1$. Intuitively, privacy when $f(P) = 0$ follows from the fact that $S_P$ reveals almost no information about the key $k$, and so the tuple $(S_P, E_{\rho,P})$ also reveals essentially no information about $\rho$ by the perfect security of the QOTP [AMTD00].

The compiler above opens an avenue towards porting results in classical secret sharing to the realm of quantum secret sharing by combining classical secret sharing schemes satisfying relevant properties (such as efficient sharing/reconstruction and small share size) with an appropriate QECC. The apparent bottleneck in this approach is the selection of the QECC, since designing efficient coding schemes for quantum states is more challenging than for classical strings. However, remarkably, we may take any code $\mathsf{QC}$ realizing *any* monotone function $f'$ satisfying $f' \geq f$. This means that even if we do not know efficient QECCs for $f$, we may instead hope to use an efficient QECC for some $f' \geq f$. Moreover, observe that we do not require any privacy properties from the *QECC* QC. In fact, the privacy of our quantum scheme follows directly from the privacy of the *classical* scheme combined with the perfect security of the QOTP, and we only need the QECC to distribute the encrypted state and overcome the no-cloning theorem.

We discuss the compiler in more detail in Section 3.

**Heavy monotone functions.** After setting up our abstract approach, we would like to instantiate it in concrete settings. Therefore, we turn our attention to the rich class of heavy monotone functions we discussed in Section 1.1 (see Definition 1). Taking into account Proposition 1 and the adjacent discussion, we know that there are heavy monotone functions $f$ computed by poly($n$)-size monotone circuits but which require exponentially large monotone span programs. This means that known methods [Smi00] for quantum secret sharing schemes realizing $f$ require exponential share size.

We use our compiler to obtain the first quantum secret sharing schemes with information ratio below 1 and also the first polynomial-time quantum secret sharing schemes for all heavy monotone functions. We discuss this in more detail in the remainder of this overview. The crucial observation behind this is that $t$-heavy monotone functions $f$ satisfy the following property: If $\mathsf{Th}_n^t$ denotes the "$t$-out-of-$n$" threshold function such that $\mathsf{Th}_n^t(P) = 1$ if and only if $|P| \geq t$, then

$$\mathsf{Th}_n^t \geq f.$$

This is useful because we know simple and highly efficient QECCs realizing $\mathsf{Th}_n^t$. For example, in this case we may take $f' = \mathsf{Th}_n^t$ and the efficient quantum Shamir threshold secret sharing scheme [CGL99] as our QECC QC in the compiler described above.

**Computational quantum secret sharing of long messages with short shares.** We exploit our compiler above to obtain Theorem 1, i.e., efficient computational quantum secret sharing schemes with constant information ratio from standard hardness assumptions.

Note that when sharing a secret composed of $m$ qubits the QOTP encryption in the compiler requires a key $k$ of length $2m$ and outputs an encrypted state composed of $m$ qubits as well. First, we may instantiate the classical computational secret sharing scheme SS as follows: Instead of sharing the key $k$ directly, replace it by the output of a post-quantum secure pseudorandom generator [BCKM21] with a much shorter uniformly random seed $s$, and share $s$ using an appropriate classical secret sharing scheme realizing $f$. Then, it remains to instantiate the QECC appropriately so as to encode the $m$ qubits into shares as short as possible. To this end, we choose an appropriate Calderbank-Shor-Steane (CSS) code [CS96, Ste96] as the QECC.

CSS codes provide a general framework for designing QECCs from *classical* linear codes[‖]. Such codes enjoy efficient encoding and decoding procedures [NC10, Section 10.4.2], and they may be seen as being analogous to packed secret sharing. We obtain our QECC by combining Reed-Solomon codes with appropriate parameters via this framework. More details can be found in Section 4.

**Efficient computational quantum secret sharing.** We now discuss how to obtain efficient computational quantum secret sharing schemes for all heavy functions in monotone P or in mNP. When $f$ is heavy and is computed by poly($n$)-size monotone circuits, we can set our classical scheme SS to be Yao's scheme [Yao89, VNS+03], whose privacy is based on the existence of one-way functions, or the Komargodski-Naor-Yogev scheme [KNY14], whose privacy is based on the

---

[‖]We say that $\mathsf{C} \subseteq \mathbb{F}_q^n$ is a *linear code* if $\mathsf{C}$ is a subspace of $\mathbb{F}_q^n$. For an extensive survey of linear codes, see [GRS22].

existence of witness encryption for NP and one-way functions. Note that both schemes are efficient. As our quantum erasure correcting code QC, we plug in quantum Shamir's scheme (Lemma 5). This leads to Theorems 2 and 3, which state that there exist efficient computational quantum secret sharing schemes for all heavy functions in monotone P or mNP, respectively, under the hardness assumptions detailed above.

For more details, see Section 5 and Section 6.

**Perfect quantum secret sharing with share size breaking the circuit size barrier.** We can also exploit the compiler to lift state-of-the-art results beyond computational secret sharing. In particular, we focus on a recent line of work improving general perfect classical secret sharing, and lift it to the quantum setting. This leads to Theorem 4, which states that there exist perfect quantum secret sharing schemes with a total share size of $1.5^{n+o(n)}$ classical bits and $O(n \log n)$ qubits realizing all heavy monotone functions.

To do this, we take any heavy monotone function $f$ and set SS to be the classical secret sharing scheme realizing $f$ constructed by Applebaum and Nir [AN21] with total share size at most $1.5^{n+o(n)}$, and QC to be the quantum Shamir secret sharing scheme for threshold functions with shares of size $O(n \log n)$ qubits [CGL99]. This leads to total share size of $1.5^{n+o(n)}$ classical bits and $O(n \log n)$ qubits, as desired. The security of the compiler in the information-theoretic setting immediately yields the desired result.

We discuss this in more detail in Section 7.

**Quantum secret sharing with multiple copies.** We also use our compiler to upper bound the number of copies of the quantum secret required to design quantum secret sharing schemes realizing *all* $t$-heavy monotone functions. More precisely, we exploit the aforementioned fact that

$$\mathsf{Th}_n^t \geq f$$

for any $t$-heavy monotone function $f$. Using this, we instantiate our compiler with an appropriate classical secret sharing scheme SS realizing $f$ (which is not bound by the no-cloning theorem), and instantiate our QECC QC with an efficient perfect quantum secret sharing scheme realizing the much simpler threshold function $\mathsf{Th}_n^t$ using $\max(1, n - 2t + 2)$ copies of the quantum secret. As a result, for any $t$-heavy monotone function $f$, we get a quantum secret sharing scheme realizing it using $\max(1, n - 2t + 2)$ copies of the secret.

We show an explicit construction of *efficient* perfect quantum secret sharing schemes realizing any $\mathsf{Th}_n^t$. Using this, we obtain Theorem 5, stating that we can construct *efficient* computational quantum secret sharing schemes realizing all $t$-heavy monotone functions $f$ in monotone P from standard hardness assumptions using $\max(1, n - 2t + 2)$ copies, by instantiating SS with Yao's scheme for $f$ [Yao89, VNS+03] and instantiating QC with the scheme we constructed for $\mathsf{Th}_n^t$. Since every monotone function is 1-heavy, setting $t = 1$ in the theorem above yields Corollary 2, which states that every monotone function in monotone P is realized by an efficient computational quantum secret sharing scheme using at most $n$ copies of the secret.

To get Theorem 6, which states that we can construct perfect quantum secret sharing schemes realizing all $t$-heavy monotone functions $f$ using $\max(1, n - 2t + 2)$ copies, we may, for example, instantiate SS with the perfect classical secret sharing scheme by Applebaum and Nir [AN21]. More details can be found in Section 8.

**Quantum secret sharing beyond heavy monotone functions.** Finally, we discuss how we can use our compiler to extend the results on quantum secret sharing above well beyond heavy monotone functions. More details can be found in Section 10.

**Weighted heavy monotone functions.** As we saw in Section 1.1, the family of weighted-heavy monotone functions (Definition 2) strictly generalizes the classes of heavy monotone functions and weighted threshold functions, as made precise in Proposition 2.

It turns out that we can apply our compiler to weighted heavy functions in a similar fashion to how we proceeded for heavy functions and generalize many of our results. The key observation that enables this, similarly to the case of heavy functions, is that if $f$ is a $(w, t)$-weighted heavy monotone function, it holds that $f \leq f'$ with $f'$ a weighted *threshold* function with the same weight function $w$ and threshold $t$. Therefore, we may instantiate our compiler with an appropriate classical secret sharing scheme SS realizing $f$ and a QECC QC for the weighted threshold function $f'$. If the sum of the weights $W = \sum_{i=1}^{n} w(i) = \text{poly}(n)$ and $t > W/2$, as is already the case in Proposition 1, we can simply take the QECC to be quantum Shamir secret sharing over $W$ parties [CGL99], and then give $w(i)$ quantum shares to the $i$-th party.

In the particularly relevant case where $f$ is computed by polynomial-size monotone circuits, we can combine the approach above with Yao's classical scheme realizing $f$ to obtain Theorem 7: There exist efficient computational quantum secret sharing schemes realizing all such weighted heavy monotone functions $f$ under standard hardness assumptions. This generalizes Theorem 2.

**Trees of heavy functions.** Similarly to the approach undertaken by Yao [Yao89, VNS+03], we can compose our quantum secret sharing schemes further to realize monotone functions computed by trees composed of gates computing weighted heavy functions. For the sake of exposition, we focus on the setting of computational privacy. Suppose that $f$ is computed by a tree $T$ whose gates compute weighted heavy functions, and we wish to share $|\psi\rangle$ according to $f$. Following [Yao89], we can start by placing $|\psi\rangle$ on the output wire. Let $g$ be the weighted heavy function on, say, $a$ input bits computing the gate to which this output wire corresponds. Then, we share $|\psi\rangle$ using a quantum secret sharing scheme realizing $g$, leading to $a$ quantum shares $S_1, \ldots, S_a$. We place the $i$-th share $S_i$ on the $i$-th in-wire of the gate computing $g$, and repeat this process with each share until we reach the leaves of the tree.

The process above yields an efficient computational quantum secret sharing scheme realizing $f$ provided that (i) We have efficient quantum secret sharing schemes for each gate in the tree, and (ii) The dimension of the quantum state to be shared in each step is always $\text{poly}(n)$. Therefore, under standard hardness assumptions, we can consider all constant-depth trees whose gates are computed by weighted heavy functions in monotone P with $\text{poly}(n)$ total weight satisfying the no-cloning property by applying the efficient computational quantum secret sharing scheme from Theorem 7 iteratively to each gate. This corresponds to Theorem 8.

## 1.3 Related work

As already discussed above, previous works on quantum secret sharing has only considered perfectly secure schemes. Furthermore, the work on schemes for general no-cloning monotone functions have mostly focused on the *existence* of such schemes [Got00, Smi00]. The share size of Gottesman's scheme [Got00] for realizing a monotone function $f$ corresponds essentially to the size of $f$ when written as a monotone formula. Smith [Smi00] constructed schemes realizing $f$ whose share size corresponds to the size of the smallest monotone span program computing $f$. In both cases, as we show here, there are many no-cloning monotone functions which require exponentially long shares under the schemes above, including some which are computed by polynomial-size monotone circuits.

Our work is the first to introduce computational privacy for quantum secret sharing schemes and the first to study notions of *efficiency*, such as polynomial-time sharing and reconstruction or small share size, for quantum secret sharing schemes realizing a broad class of monotone functions. Exploiting standard computational hardness assumptions to obtain significantly more efficient quantum secret sharing schemes, sometimes beyond what is possible in the information-theoretic setting, had not been done prior to this work, although this has been standard in the classical setting since the early work of Yao [Yao89, VNS+03] and Krawczyk [Kra94].

The study of the share size of classical secret sharing schemes already makes an appearance in the work of Benaloh and Leichter [BL88], which shows how to design general secret sharing schemes among $n$ parties with shares of size $O(2^n/\sqrt{n})$. This remained the state-of-the-art result until a recent line of work [LVW18, LV18, ABF+19, ABNP20, AN21] managed to reduce the share size over arbitrary monotone functions to $1.5^{n+o(n)}$. Remarkably, the existence of a monotone function requiring shares of size $\Omega(\frac{nm}{\log n})$, where $n$ is the number of parties and $m$ is the length of the secret, obtained by Csirmaz [Csi97] remains the state of the art lower bound. In contrast, for the special case of classical linear secret sharing schemes we know that shares must be exponentially long even if we only wish to share 1 bit [RPRC16, PR17]. Gottesman [Got00] proved the only known lower bound on the share size of quantum secret sharing schemes, which states that the dimension of each important share (i.e., a share which contributes to reconstruction) must be at least as large as the dimension of the secret state. However, if one only wishes to share a *classical* secret with a quantum secret sharing scheme, then sharing $2n$ bits requires quantum shares composed by at least $n$ qubits, and this is tight [Got00].

Other important variants of secret sharing, such as weak and verifiable secret sharing, have also been extended to the quantum setting for threshold functions [CGS05, BCG+06]. Such variants have proved useful in the design of quantum multiparty computation protocols. In particular, note that [CGS05] uses a similar compiler to ours in a different context and for unrelated goals. There has also been prior work on optimizing aspects of quantum secret sharing incomparable to those considered here. Some works have attempted to minimize the number and size of the quantum shares at the expense of larger classical components [NMI01, FG12], while others have considered the case where a subset of the parties is restricted to be classical [LQM13]. Finally, we note that the high level idea of hybrid encoding has also been utilized for different problems, such as for secure multiparty computation [BCG+06] and for fully homomorphic encryption [BJ15].

## 2 Preliminaries

### 2.1 Notation

We denote sets by uppercase letters such as $S$ and $T$, and denote $\{1, \ldots, n\}$ as $[n]$. For a vector $v$ and a set $S$, we write $v_S$ to denote $(v_i)_{i \in S}$. For a family of sets $\{A_i\}_{i \in [n]}$ and a set $S \subseteq [n]$, we let $A_S = \bigtimes_{j \in S} A_j$. For a set $\mathcal{R}$, we write $R \leftarrow \mathcal{R}$ to indicate that $R$ is uniformly distributed on $\mathcal{R}$. We use $\dim(\mathcal{H})$ to denote the dimension of a Hilbert space $\mathcal{H}$. We denote the base-2 logarithm by log. With a slight overloading of notation, we will also use $\rho \in \mathcal{H}$ to denote a density matrix $\rho$ acting on $\mathcal{H}$. Whether we mean a vector in $\mathcal{H}$ (representing a pure state) or a density matrix acting on $\mathcal{H}$ (representing a mixed state) will be clear from context and variable name (such as $|\psi\rangle$ versus $\rho$). We use monotone P to denote the set of functions that have a polynomial size *monotone circuit*. We use $\lambda$ to denote a security parameter.

### 2.2 Quantum information theory

In this section, we give an overview of the quantum information theory concepts we will be using throughout the paper.

**Definition 3** (Total variation distance). *The* total variation distance *between two random variables $X, Y$ supported on the same set $\mathcal{R}$ is defined as*

$$\Delta(X, Y) = \max_{\mathcal{A} \subseteq \mathcal{R}} |\Pr[X \in \mathcal{A}] - \Pr[Y \in \mathcal{A}]| = \frac{1}{2} \sum_{a \in \mathcal{R}} |\Pr[X = a] - \Pr[Y = a]|.$$

The analogue of the total variation distance in the quantum setting is the *trace distance*.

**Definition 4** (Trace distance [NC10]). *The* trace distance *between two density matrices $\rho$ and $\sigma$ with the same dimensions is defined as*

$$D(\rho, \sigma) = \frac{1}{2} \operatorname{tr} |\rho - \sigma|.$$

We also define the computational analogue, advantage pseudometric $A$, which satisfies similar properties as the trace distance. We later use it to prove lifting of computational privacy.

**Definition 5** (Advantage pseudometric). *For a family $\mathcal{F}$ of quantum circuits with single bit classical output and for any two density matrices $\rho, \sigma$ of appropriate dimension, the* advantage *of $\mathcal{F}$ for distinguishing $\rho$ versus $\sigma$ is defined as*

$$A_{\mathcal{F}}(\rho, \sigma) = \max_{C \in \mathcal{F}} |\Pr[C(\rho) = 1] - \Pr[C(\sigma) = 1]|.$$

See Appendix A for an overview of useful properties of trace distance and advantage.

**Quantum one-time pad encryption.** We recall quantum one-time pad encryption (QOTP) [AMTD00], which perfectly hides any quantum message using a random classical key. While we describe it for qubits, the generalization to qudits is straightforward.

The quantum one-time pad encryption scheme is defined by a pair of quantum encryption and decryption circuits (OTPEnc, OTPDec) with $\mathsf{OTPEnc} : (\mathbb{C}^2)^{\otimes n} \times \{0,1\}^{2n} \to (\mathbb{C}^2)^{\otimes n}$ and $\mathsf{OTPDec} : (\mathbb{C}^2)^{\otimes n} \times \{0,1\}^{2n} \to (\mathbb{C}^2)^{\otimes n}$ defined as

$$\mathsf{OTPEnc}(\rho, k) = (X_1^{k_1} Z_1^{k_2} \otimes \cdots \otimes X_i^{k_{2i-1}} Z_i^{k_{2i}} \otimes \cdots \otimes X_n^{k_{2n-1}} Z_n^{k_{2n}})(\rho),$$

$$\mathsf{OTPDec}(\rho, k) = (Z_1^{k_1} X_1^{k_2} \otimes \cdots \otimes Z_i^{k_{2i-1}} X_i^{k_{2i}} \otimes \cdots \otimes Z_n^{k_{2n-1}} X_n^{k_{2n}})(\rho),$$

for any message $\rho \in (\mathbb{C}^2)^{\otimes n}$ and key $k \in \{0,1\}^{\otimes 2n}$, where $X_i, Z_i$ represent the quantum operation applying the standard Pauli gates $X, Z$ respectively to the $i$-th qubit.

**Lemma 1** ([AMTD00]). *The quantum one-time pad encryption scheme is correct and perfectly secure for a randomly chosen key. That is,*

$$\mathsf{OTPDec}(\mathsf{OTPEnc}(\rho, k), k) = \rho$$

*for any key $k \in \{0,1\}^{2n}$, and*

$$\sum_{k \in \{0,1\}^{2n}} \frac{1}{2^{2n}} \mathsf{OTPEnc}(\rho, k) = \sum_{k \in \{0,1\}^{2n}} \frac{1}{2^{2n}} \mathsf{OTPEnc}(\sigma, k)$$

*for any two quantum states $\rho, \sigma \in (\mathbb{C}^2)^{\otimes n}$.*

## 2.3 Quantum adversarial model

We now introduce our quantum adversarial model. By a *QPT adversary* or circuit $C$, we mean a non-uniform family of circuits $\{C_\lambda\}_{\lambda \in \mathbb{Z}^+}$ with 1-bit classical output where each circuit has size bounded by $\text{poly}(\lambda)$ and is allowed to only use a fixed basis set of gates (e.g., $\{\mathsf{H}, \mathsf{S}, \mathsf{CNOT}, \mathsf{T}\}$), ancilla qubits each initialized to $|0\rangle$ and measurements only in standard computational basis. Furthermore, unless we explicitly state otherwise, we will assume that the adversary has access to quantum advice: that is, $C_\lambda$ in addition to its input also gets a $\text{poly}(\lambda)$ size quantum state $\rho_\lambda$ that depends only (but non-uniformly) on $\lambda$. When we say that some cryptographic scheme is *post-quantum secure*, we will mean that it is secure against QPT adversaries.

Note that models both with and without quantum advice have been considered in the literature. For example, Watrous [Wat09] and Bartusek, Coladangelo, Khurana, and Ma [BCKM21] consider the quantum advice model, while Adcock and Cleve [AC02] consider a model without advice. In the case of decision problems, it is not known if the quantum advice model is strictly stronger [Aar18].

## 2.4 Classical secret sharing

We now introduce a definition of classical secret sharing, which allows a party to distribute a classical secret among $n$ parties so that only certain subsets of parties are allowed to recover it. The definition takes into account a monotone function $f$, indicating which sets of parties are then authorized to recover the secret, and which sets do not obtain information about the secret.

**Definition 6** (Classical secret sharing [Bei11]). *Fix a number of parties $n \in \mathbb{Z}^+$, a randomness domain $\mathcal{R}$, a secret domain $S$, and share domains $S_1, \ldots, S_n$. A classical secret sharing scheme with perfect privacy realizing the monotone function $f : \{0,1\}^n \to \{0,1\}$ is a tuple of functions*

$$\mathsf{SS} = (\mathsf{Share}, (\mathsf{Rec}_P)_{P \subseteq [n]})$$

*where $\mathsf{Share} : S \times \mathcal{R} \to S_{[n]}$ and $\mathsf{Rec}_P : S_P \to S$ are deterministic functions satisfying the following properties for all $P \subseteq [n]$:*

- **Correctness:** *If $f(P) = 1$, then for all $s \in S$ it holds that*

$$\Pr_{R \leftarrow \mathcal{R}}[\mathsf{Rec}_P(\mathsf{Share}(s; R)_P) = s] = 1.$$

- **Perfect privacy:** *If $f(P) = 0$, then for all secrets $a, b \in S$ and share vectors $v \in S_P$ we have*

$$\Pr_{R \leftarrow \mathcal{R}}[\mathsf{Share}(a; R)_P = v] = \Pr_{R \leftarrow \mathcal{R}}[\mathsf{Share}(b; R)_P = v].$$

When constructing secret sharing schemes, there are several parameters and properties of interest. First and foremost, we would like to ensure that the sharing and reconstruction procedures run in time polynomial in the number of parties $n$. We call schemes with this property *efficient*. Another natural and well studied measure of complexity is the *size* of the shares in a secret sharing scheme.

**Definition 7** (Share size). *Given a secret sharing scheme $\mathsf{SS}$ over the share domains $S_1, \ldots, S_n$, we define its* share size, *denoted by $\mathsf{size}(\mathsf{SS})$, as*

$$\mathsf{size}(\mathsf{SS}) = \sum_{i=1}^{n} \lceil \log |S_i| \rceil.$$

We will be interested in secret sharing schemes with several different privacy guarantees. We can replace the perfect privacy requirement in Definition 6 with weaker, but still natural, requirements to obtain statistical and computational secret sharing. For the latter, we introduce a security parameter that we pass to the scheme.

**Definition 8** (Statistical privacy for classical secrets). *We say that a secret sharing scheme SS realizing a monotone function $f$ is $\varepsilon$-statistically private if for all $P \subseteq [n]$ such that $f(P) = 0$ and secrets $a, b \in S$ it holds that*

$$\Delta(\mathsf{Share}(a; R_1)_P, \mathsf{Share}(b; R_2)_P) \leq \varepsilon,$$

*where $R_1 \leftarrow \mathcal{R}$ and $R_2 \leftarrow \mathcal{R}$ are independent random variables.*

**Definition 9** (Post-quantum computational privacy for classical secrets). *We say that a secret sharing scheme SS realizing a monotone function $f$ is* post-quantum computationally-private, *or simply* post-quantum computational, *if for all $P \subseteq [n]$ such that $f(P) = 0$, all secrets $a, b \in S$, and for any QPT adversary $\{C_\lambda\}_\lambda$, we have*

$$\left| \Pr_{R \leftarrow \mathcal{R}}[C_\lambda(\mathsf{Share}(a; 1^\lambda, R)_P) = 1] - \Pr_{R \leftarrow \mathcal{R}}[C_\lambda(\mathsf{Share}(b; 1^\lambda, R)_P) = 1] \right| \leq \mathsf{negl}(\lambda).$$

For brevity, we will hide the security parameter and the random coins $R$ of the share functions when we do not need to use them explicitly.

## 2.5 Quantum erasure-correcting codes

We introduce the definition of a quantum erasure correcting code (QECC) [GBP97], which allows to encode a quantum state into another quantum state of larger dimension, so that the original one can be retrieved perfectly even when there are erasures (arbitrary errors at known positions).

**Definition 10** (Quantum erasure correcting code). *We say a pair of trace-preserving quantum operations $\mathsf{QC} = (\mathsf{QC.Enc}, \mathsf{QC.Dec})$ is a* quantum erasure correcting code (QECC) *over the input space $\mathcal{H}_{inp}$ and output space $\mathcal{H}_{out} = \bigotimes_{i \in [n]} \mathcal{H}_i$ for $P \subseteq [n]$ if for any quantum operation $\Lambda$ on $\mathcal{H}_{out}$ that acts as the identity on $\mathcal{H}_i$ for all $i \in P$, it holds for all states $\rho$ on $\mathcal{H}_{inp}$ that*

$$(\mathsf{QC.Dec} \circ \Lambda \circ \mathsf{QC.Enc})(\rho) = \rho \otimes \sigma$$

*for some state $\sigma$.*

*If $(\mathsf{QC.Enc}, \mathsf{QC.Dec}_P)$ is a QECC for all sets $P \subseteq [n]$ such that $f(P) = 1$ for a monotone function $f : \{0,1\}^n \to \{0,1\}$, then we say that the family of functions $(\mathsf{QC.Enc}, (\mathsf{QC.Dec}_P)_{P \subseteq [n]})$ is a QECC realizing $f$. As a shorthand, we define $\mathsf{QC.Rec}_P(\tau) = \mathsf{QC.Dec}(\tau \otimes (|0\rangle\langle 0|)^{\otimes \overline{P}})$. A quantum code that encodes $k$ $q$-ary qudits into $n$ $q$-ary qudits and can correct any $d - 1$ erasures is said to be an $[[n, k, d]]_q$ code.*

## 2.6 Quantum secret sharing

A natural analogue of classical secret sharing is sharing quantum states, introduced by [HBB99, CGL99, KKI99]. We start by formally defining quantum secret sharing with perfect privacy, and then introduce for the first time the alternative notion of computational privacy for quantum secret sharing.

**Definition 11** (No-cloning function). *A monotone function $f : \{0,1\}^n \to \{0,1\}$ is called* no-cloning *if we have $f(\overline{P}) = 0$ for any $P \subseteq [n]$ with $f(P) = 1$.*

**Definition 12** (Quantum secret sharing). *Fix a number of parties $n \in \mathbb{Z}^+$, a Hilbert space $\mathcal{S}$ for the secret, and Hilbert spaces $\mathcal{H}_1, \ldots, \mathcal{H}_n$ for the shares. Let $f : \{0,1\}^n \to \{0,1\}$ be a no-cloning monotone function. A* quantum secret sharing (QSS) scheme with perfect privacy realizing $f$ *is a tuple of trace-preserving quantum operations*

$$\mathsf{QSS} = (\mathsf{Share}, (\mathsf{Rec}_P)_{P \subseteq [n]})$$

*that satisfy the following properties for all $P \subseteq [n]$:*

- **Correctness:** *If $f(P) = 1$, then $(\mathsf{Share}, \mathsf{Rec}_P)$ is a QECC for $P$.*

- **Perfect Privacy:** *If $f(P) = 0$, then for any $|\psi_1\rangle, |\psi_2\rangle \in \mathcal{S}$ it holds that*

$$\mathrm{tr}_{\overline{P}}(\mathsf{Share}(|\psi_1\rangle\langle\psi_1|)) = \mathrm{tr}_{\overline{P}}(\mathsf{Share}(|\psi_2\rangle\langle\psi_2|)).$$

We call a scheme $\mathsf{QSS}$ *efficient* if $\mathsf{QSS.Share}, \mathsf{QSS.Rec}$ are polynomial size circuits. Note that, in particular, efficient schemes have polynomial size shares. We can define weaker notions of privacy analogously to classical secret sharing in Section 2.4.

**Definition 13** (Statistical privacy for quantum secrets). *We say that a quantum secret sharing scheme $\mathsf{QSS}$ realizing $f$ is $\varepsilon$-statistically private if for all $P \subseteq [n]$ such that $f(P) = 0$ and any secrets $|\psi_1\rangle, |\psi_2\rangle \in \mathcal{S}$ it holds that*

$$D(\mathrm{tr}_{\overline{P}}(\mathsf{Share}(|\psi_1\rangle\langle\psi_1|)), \mathrm{tr}_{\overline{P}}(\mathsf{Share}(|\psi_2\rangle\langle\psi_2|))) \leq \varepsilon.$$

Observe that perfect privacy corresponds to 0-statistical privacy.

As in the classical case, we have two different notions of computational privacy, namely, against quantum adversaries with no advice and quantum adversaries with quantum advice.

**Definition 14** (Computational privacy for quantum secrets). *We say that a quantum secret sharing scheme $\mathsf{QSS}$ realizing $f$ is* computationally-private, *or simply* computational, *if for all $P \subseteq [n]$ such that $f(P) = 0$, any secrets $|\psi_1\rangle, |\psi_2\rangle \in \mathcal{S}$, and any QPT adversary $\{C_\lambda\}_\lambda$ we have*

$$\left| \Pr\left[ C_\lambda(\mathrm{tr}_{\overline{P}}(\mathsf{Share}(|\psi_1\rangle\langle\psi_1|; 1^\lambda))) = 1 \right] - \Pr\left[ C_\lambda(\mathrm{tr}_{\overline{P}}(\mathsf{Share}(|\psi_2\rangle\langle\psi_2|; 1^\lambda))) = 1 \right] \right| \leq \mathsf{negl}(\lambda).$$

Note that requiring privacy (similarly, correctness) for pure states is sufficient, and privacy (correctness) for mixed states of any polynomial size follows by a simple diagonalization and triangle inequality argument. Also observe that any quantum secret sharing scheme for $f$ is also a quantum erasure correcting code realizing $f$.

## 3 The compiler

In this section, we present a simple compiler that allows us to obtain several new results results by lifting a wide range of results on classical secret sharing to the quantum setting. As discussed in Section 1.2, these include:

- Efficient computational quantum secret sharing schemes for heavy functions in monotone P with information ratio *well below* 1 (i.e., with shares much shorter than the secret) from standard hardness assumptions. This is impossible in the information-theoretic setting.

- Efficient computationally-private quantum secret sharing for all heavy functions in monotone P and mNP. In contrast, known schemes [Got00, Smi00] require shares of exponential size for these functions (see Proposition 1 and adjacent discussion). The techniques can be further applied to classes of monotone functions that generalize heavy functions, including weighted heavy functions and trees of weighted heavy functions.

- Similar techniques can also be applied to obtain perfect quantum secret sharing schemes for heavy functions with share size $1.5^{n+o(n)}$, breaking the circuit size barrier.

- Polynomial size computational schemes for any function in monotone P given sufficiently many copies of the secret.

We now move on to the compiler. Unless otherwise specified, we assume any state space $\mathcal{H}$ is $(\mathbb{C}^2)^{\otimes \ell}$ for some $\ell \in \mathbb{Z}^+$, that is, we are working with qubits. The generalization to qudits is straightforward. We will also assume $\ell = 1$ unless otherwise stated.

**Compiler Description.** The compiler combines a classical secret sharing scheme SS realizing a no-cloning monotone function $f$, and a quantum error correcting code QC realizing an appropriate no-cloning monotone function $f' \geq f$, to create a quantum secret sharing scheme QSS realizing $f$. In order to secret share a quantum state $\rho$, the scheme QSS.Share first samples a random classical key k and computes the encryption $\rho'$ of $\rho$ using the quantum one-time pad. We then distribute the classical key k using SS and the quantum state $\rho'$ using QC. Intuitively, the privacy of the overall scheme follows directly from the privacy of the classical scheme SS and privacy of the quantum one-time pad.

The reconstruction procedure QSS.Rec is straightforward. We simply let the set of parties $P$ reconstruct the state $\rho'$ using the decoding procedure for QC and the key k using the reconstruction procedure of SS. The quantum secret $\rho$ is then reconstructed by decrypting $\rho'$ with the obtained key k via the quantum one-time pad.

**QSS Share for $f$: QSS.Share($\rho$).**

1. Sample key $k \leftarrow \{0,1\}^{2 \log \dim \mathcal{S}}$.
2. Compute $\rho' = \mathsf{OTPEnc}(\rho, k)$, the encryption of $\rho$ using QOTP with key k.
3. Let $(E_1, \ldots, E_n) = \mathsf{QC.Enc}(\rho')$ be the encoding of $\rho'$.
4. Let $(S_1, \ldots, S_n) = \mathsf{SS.Share}(k)$ be a sharing of k.
5. Set $(S_i, E_i)$ as the share for party $P_i$.

**QSS Reconstruct for $f$: QSS.Rec$_P((S_i, E_i)_{i \in P})$.**

1. Compute $\rho' = \mathsf{QC.Rec}_P((E_i)_{i \in P})$.
2. Compute $k = \mathsf{SS.Rec}_P((S_i)_{i \in P})$.
3. Compute $\rho = \mathsf{OTPDec}(\rho', k)$.
4. Output $\rho$.

We now formally state the main theorem.

**Theorem 9** (QSS Compiler). *Let $f, f' : \{0,1\}^n \to \{0,1\}$ be no-cloning monotone functions such that $f' \geq f$. Let $\mathsf{QC} = (\mathsf{QC.Enc}, (\mathsf{QSS.Rec}_P)_{P \subseteq [n]})$ be a QECC realizing $f'$ and $\mathsf{SS} = (\mathsf{SS.Share}, (\mathsf{SS.Rec}_P)_{P \subseteq [n]})$ be a [post-quantum computational, statistical, perfect] classical secret sharing scheme realizing $f$. Then, $\mathsf{QSS}$ is a [computational, statistical, perfect] quantum secret sharing scheme for $f$ with total share size*

$$\mathsf{size}(\mathsf{QC}) + 2\log(\dim \mathcal{S}) \cdot \mathsf{size}(\mathsf{SS}).$$

*Moreover, $\mathsf{QSS}$ has efficient sharing and reconstruction procedures whenever $\mathsf{QC}$ and $\mathsf{SS}$ do.*

*Proof.* Let us denote by $\xi_{\rho,\mathsf{k}} = \mathsf{QC.Enc}(\mathsf{OTPEnc}(\rho, \mathsf{k}))$ the encoding of the state $\mathsf{OTPEnc}(\rho, \mathsf{k})$ using $\mathsf{QC}$, and $\tau_{\mathsf{k},r} = |\mathsf{SS.Share}(\mathsf{k}, r)\rangle\langle\mathsf{SS.Share}(\mathsf{k}, r)|$ a sharing of the key $\mathsf{k}$ when the random input is $r$. Then, the scheme $\mathsf{QSS}$ can be formally described as

$$\mathsf{QSS.Share}(\rho) = \sum_{\mathsf{k} \in \{0,1\}^{2\log \dim \mathcal{S}}} \sum_{r \in \mathcal{R}} \frac{1}{2^{2\log \dim \mathcal{S}}} \frac{1}{|\mathcal{R}|} (\xi_{\rho,\mathsf{k}} \otimes \tau_{\mathsf{k},r}),$$

$$\mathsf{QSS.Rec}_P(\sigma) = \mathsf{OTPDec}(\mathsf{QC.Rec}_P(\mathrm{tr}_{key}(\sigma))), \mathsf{SS.Rec}_P(\mathrm{tr}_{state}(\sigma))),$$

where $\mathrm{tr}_{key}, \mathrm{tr}_{state}$ denotes tracing out the subsystem corresponding to the shares of the key and shares of the quantum secret respectively. Here in the formalism, for simplicity, we encode the shares of the classical keys as qubits in basis states, however, they can be kept as classical shares in practice without any change to the scheme.

Showing correctness is straightforward. Lifting the perfect privacy of $\mathsf{SS}$ to perfect privacy of $\mathsf{QSS}$ will be a simple corollary of the statistical privacy lifting proven below. However, as a warm-up we first prove it directly. Consider any $P \subseteq [n]$ with $f(P) = 0$. Then,

$$\mathrm{tr}_{\overline{P}}(\mathsf{QSS.Share}(\rho)) = \mathrm{tr}_{\overline{P}}\left(\sum_{\mathsf{k} \in \{0,1\}^{2\log\dim\mathcal{S}}} \sum_{r \in \mathcal{R}} \frac{1}{2^{2\log\dim\mathcal{S}}} \frac{1}{|\mathcal{R}|} \xi_{\rho,\mathsf{k}} \otimes \tau_{\mathsf{k},r}\right)$$

$$= \sum_{\mathsf{k} \in \{0,1\}^{2\log\dim\mathcal{S}}} \mathrm{tr}_{\overline{P}}\left(\frac{1}{2^{2\log\dim\mathcal{S}}} \xi_{\rho,\mathsf{k}}\right) \otimes \mathrm{tr}_{\overline{P}}\left(\sum_{r \in \mathcal{R}} \frac{1}{|\mathcal{R}|} \tau_{\mathsf{k},r}\right)$$

$$= \sum_{\mathsf{k} \in \{0,1\}^{2\log\dim\mathcal{S}}} \mathrm{tr}_{\overline{P}}\left(\frac{1}{2^{2\log\dim\mathcal{S}}} \xi_{\rho,\mathsf{k}}\right) \otimes \sigma_P \qquad (1)$$

$$= \mathrm{tr}_{\overline{P}}\left(\mathsf{QC.Enc}\left(\sum_{\mathsf{k} \in \{0,1\}^{2\log\dim\mathcal{S}}} \frac{1}{2^{2\log\dim\mathcal{S}}} \mathsf{OTPEnc}(\rho, k)\right)\right) \otimes \sigma_P$$

$$= \mathrm{tr}_{\overline{P}}\left(\mathsf{QC.Enc}(\sigma'_P)\right) \otimes \sigma_P \qquad (2)$$

where $\sigma_P, \sigma'_P$ are some states that depend only on $P \subseteq [n]$. Therefore, shares of an unauthorized set $P$ do not depend on the secret $\rho$. The equality in Equation (1) is due to perfect privacy of $\mathsf{SS}$ and Equation (2) is due to perfect privacy of quantum one-time pad.

Now, we show that if $\mathsf{SS}$ is $\varepsilon$-statistically private, then $\mathsf{QSS}$ is $2\varepsilon$-statistically private. First, observe the following relation between trace distance and total variation distance. Consider any two keys, $k, k' \in \{0,1\}^{2\log\dim\mathcal{S}}$. For each $v \in V = S_P$, define $p_v = \Pr_{r \leftarrow R}[\mathsf{Share}(k;r)_P = v]$ and define $p'_v$ analogously for $k'$. Then, using the fact that

$$\mathrm{tr}_{\overline{P}}\left(\sum_{r \in \mathcal{R}} \frac{1}{|\mathcal{R}|} \tau_{\mathsf{k},r}\right) = \sum_{v \in V} p_v |v\rangle\langle v|,$$

it is easy to see that

$$D\left(\mathrm{tr}_{\overline{P}}\left(\sum_{r\in\mathcal{R}}\frac{1}{|\mathcal{R}|}\tau_{\mathsf{k},r}\right),\mathrm{tr}_{\overline{P}}\left(\sum_{r\in\mathcal{R}}\frac{1}{|\mathcal{R}|}\tau_{\mathsf{k}',r}\right)\right)=D\left(\sum_{v\in V}p_v\left|v\right\rangle\!\left\langle v\right|,\sum_{v\in V}p_v'\left|v\right\rangle\!\left\langle v\right|\right)$$
$$\leq \Delta(p_v, p_v'). \tag{3}$$

We now study the privacy of QSS. Consider a set $P\subseteq [n]$ such that $f(P)=0$ and any two secret states $\rho, \rho'$. We will use a hybrid argument. First, we will argue that when we replace the shares of the key with shares of an independent uniformly distributed key, the composite shares of $\overline{P}$ for the two secrets $\rho, \rho'$ will be perfectly indistinguishable due to the perfect privacy of one-time pad from Lemma 1. That is, we will define the sharing of a random key

$$\kappa = \sum_{\substack{k'\in\{0,1\}^{2\log\dim\mathcal{S}}\\r\in\mathcal{R}}}\frac{1}{2^{2\log\dim\mathcal{S}}|\mathcal{R}|}\tau_{\mathsf{k}',r}$$

and the hybrids

$$\zeta_1 = \mathrm{tr}_{\overline{P}}\left(\sum_{k\in\{0,1\}^{2\log\dim\mathcal{S}}}\left(\frac{1}{2^{2\log\dim\mathcal{S}}}\xi_{\rho,\mathsf{k}}\right)\otimes\kappa\right),$$

$$\zeta_2 = \mathrm{tr}_{\overline{P}}\left(\sum_{k\in\{0,1\}^{2\log\dim\mathcal{S}}}\left(\frac{1}{2^{2\log\dim\mathcal{S}}}\xi_{\rho',\mathsf{k}}\right)\otimes\kappa\right),$$

and will show that

$$D(\zeta_1,\zeta_2) = 0. \tag{4}$$

Then, we will show that composite shares of $\overline{P}$ for the same secret are close in trace distance when again the shares of the key are replaced with shares of a random key, versus when they are not replaced. That is, we will show that

$$D(\mathrm{tr}_{\overline{P}}(\mathsf{QSS.Share}(\rho)),\zeta_1)\leq\varepsilon\quad\text{and}\quad D(\zeta_2,\mathrm{tr}_{\overline{P}}(\mathsf{QSS.Share}(\rho')))\leq\varepsilon. \tag{5}$$

Finally applying the triangle inequality will yield the desired result.

We start with Equation (4). By distributing the partial trace and using Lemma 11 (Item v.), we get

$$D(\zeta_1,\zeta_2)$$
$$= D\left(\mathrm{tr}_{\overline{P}}\left(\sum_{k\in\{0,1\}^{2\log\dim\mathcal{S}}}\left(\frac{1}{2^{2\log\dim\mathcal{S}}}\xi_{\rho,\mathsf{k}}\right)\right),\mathrm{tr}_{\overline{P}}\left(\sum_{k\in\{0,1\}^{2\log\dim\mathcal{S}}}\left(\frac{1}{2^{2\log\dim\mathcal{S}}}\xi_{\rho',\mathsf{k}}\right)\right)\right).$$

Then, since the quantum one-time pad perfectly hides the input when the key is uniform (see Lemma 1), we get for some state $\iota$ that

$$D(\zeta_1,\zeta_2) = D(\mathrm{tr}_{\overline{P}}(\mathsf{QC.Enc}(\iota)),\mathrm{tr}_{\overline{P}}(\mathsf{QC.Enc}(\iota))) = 0.$$

The inequalities in Equation (5) are proven using the privacy of SS and again properties of trace distance, along with Equation (3). By Lemma 11 (Items iii. and v.), we get

$$D(\zeta_1,\mathrm{tr}_{\overline{P}}(\mathsf{QSS.Share}(\rho))) = D\left(\sum_{k\in\{0,1\}^{2\log\dim\mathcal{S}}}\frac{1}{2^{2\log\dim\mathcal{S}}}\mathrm{tr}_{\overline{P}}(\xi_{\rho,\mathsf{k}})\otimes\mathrm{tr}_{\overline{P}}(\kappa),$$

$$\sum_{k\in\{0,1\}^{2\log\dim\mathcal{S}}} \frac{1}{2^{2\log\dim\mathcal{S}}} \operatorname{tr}_{\overline{P}}(\xi_{\rho,k}) \otimes \operatorname{tr}_{\overline{P}}\left(\sum_{r\in\mathcal{R}} \frac{1}{|\mathcal{R}|}\tau_{k,r}\right)$$

$$\leq \sum_{k\in\{0,1\}^{2\log\dim\mathcal{S}}} \frac{1}{2^{2\log\dim\mathcal{S}}} D\left(\operatorname{tr}_{\overline{P}}(\kappa), \operatorname{tr}_{\overline{P}}\left(\sum_{r\in\mathcal{R}} \frac{1}{|\mathcal{R}|}\tau_{k,r}\right)\right).$$

Then, using Lemma 11 (Item iv.), we get

$$D(\zeta_1, \operatorname{tr}_{\overline{P}}(\mathsf{QSS.Share}(\rho)))$$

$$\leq \sum_{k,k'\in\{0,1\}^{2\log\dim\mathcal{S}}} \frac{1}{4^{2\log\dim\mathcal{S}}} D\left(\operatorname{tr}_{\overline{P}}\left(\sum_{r\in\mathcal{R}} \frac{1}{|\mathcal{R}|}\tau_{k',r}\right), \operatorname{tr}_{\overline{P}}\left(\sum_{r\in\mathcal{R}} \frac{1}{|\mathcal{R}|}\tau_{k,r}\right)\right).$$

Observe that this is basically the statistical distance between classical sharings of keys $k, k'$. Therefore, invoking the $\varepsilon$-statistical privacy of $\mathsf{SS}$ and Equation (3), we conclude that

$$D(\zeta_1, \operatorname{tr}_{\overline{P}}(\mathsf{QSS.Share}(\rho))) \leq \varepsilon.$$

The same argument also shows that

$$D(\zeta_2, \operatorname{tr}_{\overline{P}}(\mathsf{QSS.Share}(\rho'))) \leq \varepsilon.$$

Finally, we combine these inequalities with the triangle inequality to obtain

$$\begin{aligned} &D(\operatorname{tr}_{\overline{P}}(\mathsf{QSS.Share}(\rho)), \operatorname{tr}_{\overline{P}}(\mathsf{QSS.Share}(\rho'))) \\ &\leq D(\operatorname{tr}_{\overline{P}}(\mathsf{QSS.Share}(\rho)), \zeta_1) + D(\zeta_1, \zeta_2) + D(\zeta_2, \operatorname{tr}_{\overline{P}}(\mathsf{QSS.Share}(\rho'))) \\ &\leq \varepsilon + 0 + \varepsilon \\ &= 2\varepsilon. \end{aligned}$$

Plugging in $\varepsilon = 0$ yields the desired result for perfect privacy, since $\Delta$ and $D$ are both metrics.

Lastly, we consider the setting of computational privacy. Here, we use the quantum advantage pseudometric $A$, which, as shown in Lemma 12, satisfies the same basic properties that we use above for trace distance. Hence, replacing trace distance with advantage in the proof above proves the lifting of computational privacy. More specifically, by Lemma 12 (Item vii.), we get the following.[**]

- Suppose that $\mathsf{SS}$ is computationally-private with respect to QPT adversaries with no advice, that $\mathsf{QC.Enc}$ can be implemented by quantum circuits of size $\mathrm{poly}(\lambda)$, and that any pure secret in the space of secrets can be approximated by a quantum circuit of size $\mathrm{poly}(\lambda)$.[††] Then, $\mathsf{QSS}$ is computationally-private with respect to QPT adversaries with no advice.

- If $\mathsf{SS}$ is secure against QPT adversaries with quantum advice and the shares of $\mathsf{QC}$ are at most $\mathrm{poly}(\lambda)$ qubits, then $\mathsf{QSS}$ is also secure against the same family of adversaries. $\square$

---

[**]Note that we need the extra assumptions below since otherwise one can construct pathological schemes and secrets so that the adversary obtains non-uniform quantum advice.

[††]This is readily true, for example, if $\log\dim(\mathcal{S}) = \Theta(1)$.

# 4 Computational quantum secret sharing of long messages with short shares

In this section we consider the problem of sharing a long secret consisting of multiple qubits and prove Theorem 1, which states that there exist efficient computational quantum secret sharing schemes for all heavy functions in monotone P with information ratio well below 1. Note that all previously known quantum secret sharing schemes are information-theoretic and require a share size that is as large as the secret, as proven by Gottesman [Got00], even for the very simple case of threshold monotone functions. We show that we can achieve a scheme with individual share sizes much shorter than the secret, with the help of computational hardness assumptions.

Our scheme follows the template of the general compiler presented in Section 3, with the following sub-protocols: (i) the key used for the one-time pad encryption is generated using a pseudo-random generator (PRG) and its seed is shared using a computational classical secret sharing scheme with short shares, and (ii) a CSS quantum erasure-correcting code with low share size.

We formally describe the share procedure QSS.Share below. The reconstruction procedure QSS.Rec$_P$ is as in Section 3, except that the key is generated by reconstructing the seed and evaluating the PRG.

**QSS Share for $f$: QSS.Share($\rho$).**

1. Sample a seed $\mathsf{x} \leftarrow \{0,1\}^{\ell(m)}$.
2. Compute $\mathsf{k} = \mathsf{PRG}(\mathsf{x})$.
3. Compute $\sigma = \mathsf{OTPEnc}(\rho, k)$, the encryption of $\rho$ using QOTP with key $\mathsf{k}$.
4. Let $(E_1, \ldots, E_n) = \mathsf{QC.Enc}(\sigma)$ be the encoding of $\sigma$.
5. Let $\mathsf{SS.Share}(\mathsf{x})$ be a sharing of $\mathsf{x}$.
6. Set $(S_i, E_i)$ as the share for party $P_i$.

**QSS Reconstruct for $f$: QSS.Rec$_P((S_i, E_i)_{i \in P})$.**

1. Compute $\sigma = \mathsf{QC.Rec}_P((E_i)_{i \in P})$.
2. Compute $\mathsf{x} = \mathsf{SS.Rec}_P((S_i)_{i \in P})$.
3. Compute $\mathsf{k} = \mathsf{PRG}(\mathsf{x})$.
4. Compute $\rho = \mathsf{OTPDec}(\sigma, \mathsf{k})$.
5. Output $\rho$.

Correctness of the scheme is straightforward. Computational privacy easily follows from the privacy of the underlying PRG and the classical secret sharing scheme, by an argument analogous to the proof of Theorem 9.

Observe that by plugging in a pseudo-random generator PRG with polynomial stretch, it is possible to get the same ratio as that of the QECC for sufficiently long messages. As a concrete example, we show below a scheme with asymptotic information ratio $\frac{32}{2t-n}$ for any $t$-heavy monotone function in monotone P. Note that previous schemes [Got00, Smi00], or a naive application of Theorem 9, both yield information ratio above 1 (in fact, exponentially or polynomially larger, respectively).

Before we prove our main result, we need to construct a suitable QECC. To that end, we need the following general template for CSS codes (for a definition of linear codes, see Definition 17).

**Lemma 2** (CSS Codes [GBP97, Theorem 6] and [Rai99]). *Let $q$ be a prime power, $C_1$ an $[n, k_1, d_1]_q$ linear code, and $C_2$ an $[n, k_2, d_2]_q$ linear code with $C_2^\perp \subseteq C_1$. Then, there exists an $[[n, k_1 + k_2 - n, \min(d_1, d_2)]]_q$ quantum code.*

With the help of this lemma, we can obtain a QECC with the required parameter trade-offs to achieve QSS with small share size.

**Lemma 3.** *For any $m, n, t$ with $n \geq t > \frac{n}{2}$, there is a $[[N, 2K - N, N - K + 1]]_{2^r}$ CSS code QC where the parameters are defined as follows: Let $N^*$ be such that $N^* \log_2(N^*) = \frac{2mn}{t - \frac{n}{2}}$, and set $c = \left\lceil \frac{N^*}{n} \right\rceil$ along with*

$$N = cn, \quad r = \lceil \log_2(N) \rceil \quad \text{and} \quad K = \left\lceil \frac{nc}{2} + \frac{1}{2} \left\lceil \frac{m}{r} \right\rceil \right\rceil.$$

*Then, for large enough $m$ we have*

$$2K - N \geq \left\lceil \frac{m}{r} \right\rceil \quad \text{and} \quad N - K \geq c(n - t).$$

*Proof.* In Lemma 2, use the same Reed-Solomon $[N, K, N - K + 1]_{2^r}$ code guaranteed by Lemma 13 as both $C_1$ and $C_2$. The inequalities follow by simple calculations using the values of $N, K$ given in the lemma statement. □

We are now ready to state the final theorem.

**Theorem 10** (Theorem 1, restated). *If $f$ is a t-heavy monotone function, with $t > n/2$, computed by monotone circuits of size $O(n^d)$, then there is an efficient computational quantum secret sharing scheme realizing $f$ with asymptotic information ratio at most $\frac{32}{2t-n}$ for secrets composed of at least $m = \Omega(n^{cd})$ qubits for a universal constant $c > 0$ based on the existence of post-quantum secure one-way functions.*

*Proof.* Suppose we are sharing a secret $s$ of $m$ qubits. We will use the compiler above for $f$ by setting $f' = \text{Th}_n^t$ and plugging in the code QC below, a post-quantum secure PRG with polynomial stretch as PRG, and Yao's efficient classical scheme from Lemma 4 realizing $f$ as SS.

Using the CSS code from Lemma 3, we can define the code QC for $m$ qubit inputs as:

1. Pad the input to $\lceil \frac{m}{r} \rceil r$ qubits, which is equivalently $\lceil \frac{m}{q} \rceil$ $2^r$-ary qudits,

2. Pad the input qudits to get $2K - N$ $2^r$-ary qudits,

3. Encode the input using the code from Lemma 3 and give output qudits $ic, ic+1, \ldots, (i+1)c-1$ to party $i$ for $i \in [n]$

It is easy to see that this is a QECC for $f' = \text{Th}_n^t$.

Observe that we are using $\frac{N}{n}$ $2^r$-ary qudits per party, or equivalently $\frac{Nr}{n}$ qubits to share a secret of $m$ qubits. Therefore, for the quantum code we get a ratio of

$$\frac{Nr}{mn} \leq \frac{32}{2t-n}.$$

To see the ratio corresponding to the classical part, we can assume that seed length of the PRG is $m^\gamma$ for some $\gamma < 1$ since PRG has a polynomial stretch. Then, for any constant $k > \frac{d}{1-\gamma}$ and for all $m \geq n^k$, the ratio of the classical part is

$$\frac{m^\gamma \cdot n^d}{m} \leq \frac{m^\gamma \cdot m^{d/k}}{m} \to 0$$

as $n \to \infty$.

Finally, by plugging these into the compiler, we get a quantum secret sharing scheme for $t$-heavy $f$ with an information ratio bounded by $\frac{32}{2t-n}$. □

## 5 Efficient quantum secret sharing for heavy functions in monotone P

In this section, we use our compiler from Section 3 to design efficient computational *quantum* secret sharing schemes for all heavy functions in monotone P. Before we proceed to the main theorem statement, we require the following results of Yao [Yao89, VNS+03] and Cleve, Gottesman, and Lo [CGL99].

**Lemma 4** ([Yao89, VNS+03]). *If $f$ is in monotone P, there exists an efficient post-quantum computational classical secret sharing scheme realizing $f$ based on the existence of post-quantum secure one-way functions.*

**Lemma 5** (Quantum Shamir secret sharing [CGL99]). *There is an efficient perfect quantum secret sharing scheme realizing $\mathsf{Th}_n^t$ for any $t > n/2$ with share size $O(n \log n)$.*

We are now ready to prove our result.

**Theorem 11** (Theorem 2, restated). *If $f$ is a heavy monotone function in monotone P, then there is an efficient computational quantum secret sharing scheme realizing $f$ based on the existence of post-quantum secure one-way functions.*

*Proof.* Fix an arbitrary heavy function $f : \{0,1\}^n \to \{0,1\}$ in monotone P. In particular, since $f$ is heavy, it holds that

$$\mathsf{Th}_n^t \geq f$$

for $t = \lfloor n/2 \rfloor + 1$. Therefore, it suffices to invoke Theorem 9 for $f$ with $f' = \mathsf{Th}_n^t$, set SS to be the classical scheme guaranteed by Lemma 4, and set QC to be the efficient quantum Shamir secret sharing scheme from Lemma 5. □

Note that, as discussed in Section 1, there are heavy functions in monotone P that require exponential size monotone span programs, and therefore previous methods can only yield exponential size schemes for such functions, whereas we are able to obtain polynomial size schemes above. See Section 9 for a more thorough discussion.

Similarly, we can lift the result of Komargodski, Naor, and Yogev [KNY14] to obtain efficient quantum secret sharing schemes for all heavy functions in mNP, based on the existence of post-quantum secure witness encryption. See Section 6 for a more detailed discussion.

# 6 Efficient quantum secret sharing schemes for heavy monotone functions in mNP

In this section, we further extend the results of Section 5 and show that, based on stronger hardness assumptions, we can obtain efficient schemes for heavy functions in mNP using the result of Komargodski, Naor, and Yogev [KNY14] in the classical.

Before we proceed to restate and prove our main result in this setting, we must clarify what is meant by secret sharing for mNP. First, we say that a monotone function $f$ is in mNP if the language $L = \{x \in \{0,1\}^n : f(x) = 1\}$ is in NP. The definition of secret sharing for functions $f \in$ mNP is similar to the definitions discussed in Section 2.6, except that we must also give the reconstruction procedure a valid polynomial-sie witness for the statement "$f(P) = 1$". We present a definition in the quantum setting. The corresponding definition in the classical setting is analogous (cf. [KNY14]).

**Definition 15** (Quantum secret sharing for mNP). *Fix a number of parties $n \in \mathbb{Z}^+$, a Hilbert space $\mathcal{S}$ for the secret, and Hilbert spaces $\mathcal{H}_1, \ldots, \mathcal{H}_n$ for the shares. Let $f : \{0,1\}^n \to \{0,1\}$ be a no-cloning monotone function recognizing a language $L \in$ mNP, and let $V$ be a polynomial-time verifier for $L$. A quantum secret sharing scheme (QSS) for mNP realizing the access function $f$ is a family of trace-preserving quantum operations $\mathsf{QSS} = (\mathsf{Share}, (\mathsf{Rec}_P)_{P \subseteq [n]})$ that satisfy the following properties for all $P \subseteq [n]$:*

- **Correctness:** *If $f(P) = 1$, then for any valid witness $w$ such that $V(P, w) = 1$, the following holds: if $\rho_P$ denotes the shares of the subset $P$ of parties, then $\mathsf{Rec}_P(\rho_P, P, w) = |\psi\rangle$.*

- **Computational privacy:** *If $f(P) = 0$, then for any $|\psi_1\rangle, |\psi_2\rangle \in \mathcal{S}$ and for any QPT adversary $\{C_\lambda\}_\lambda$, we have*

$$\left| \Pr\left[ C(\mathrm{tr}_{\overline{P}}(\mathsf{Share}(|\psi_1\rangle\langle\psi_1|; 1^\lambda))) = 1 \right] - \Pr\left[ C(\mathrm{tr}_{\overline{P}}(\mathsf{Share}(|\psi_2\rangle\langle\psi_2|; 1^\lambda))) = 1 \right] \right| \leq \mathsf{negl}(\lambda).$$

We will also need the following result from [KNY14].

**Lemma 6** ([KNY14]). *If $f \in$ mNP, there is an efficient post-quantum computational classical secret sharing scheme realizing $f$ based on the existence of post-quantum secure witness encryption for NP and one-way functions.*

We are now ready to prove our theorem.

**Theorem 12** (Theorem 3, restated). *For any heavy function $f : \{0,1\}^n \to \{0,1\}$ in mNP, there is a computational quantum secret sharing scheme QSS realizing $f$ with $\mathsf{size}(\mathsf{QSS}) \leq \mathrm{poly}(n)$, based on the existence of post-quantum secure witness encryption for NP and one-way functions.*

*Proof.* The argument is analogous to the proof we gave in Section 5 for Theorem 2. Namely, we invoke our compiler (Theorem 9) for $f$ with $f' = \mathsf{Th}_n^{\lfloor \frac{n}{2} \rfloor + 1}$, and set SS to be the scheme given by Lemma 6 and QC to the quantum Shamir secret sharing scheme from Lemma 5. □

# 7 Perfect quantum secret sharing with share size breaking the circuit size barrier

In this section, we use our compiler to break the circuit size barrier for the share size of perfect quantum secret sharing schemes for general heavy monotone functions, yielding Theorem 4. To prove this result, we need the following state-of-the-art for general perfect classical secret sharing schemes by Applebaum and Nir [AN21].

**Lemma 7** ([AN21]). *Given any monotone function $f : \{0,1\}^n \to \{0,1\}$, there is a perfect classical secret sharing scheme realizing $f$ with share size $1.5^{n+o(n)}$.*

We are now ready to prove the main result, which we restate below.

**Theorem 13** (Theorem 4, restated). *If $f : \{0,1\}^n \to \{0,1\}$ is a heavy monotone function, then there is a perfect quantum secret sharing scheme realizing $f$ with a total share size of $1.5^{n+o(n)}$ classical bits and $O(n \log n)$ qubits.*

*Proof.* We invoke Theorem 9 for $f$ with $f' = \mathsf{Th}_n^t \geq f$ for $t = \lfloor n/2 \rfloor + 1$, set SS to be the classical secret sharing scheme guaranteed by Lemma 7, and let QC be the quantum Shamir secret sharing scheme from Lemma 5. The number of bits and qubits in the shares follows directly from these lemmas. □

We remark that, via our compiler, any future result which improves on Lemma 7 in the classical setting immediately translates into an improved matching result in the quantum setting. As discussed in Section 1, Theorem 4 improves on previous quantum secret sharing schemes [Got00, Smi00] for heavy monotone functions, since the share size of general linear classical schemes is much larger than that guaranteed by Lemma 7, and this holds in particular for heavy monotone functions by Proposition 1.

## 8 Quantum secret sharing using multiple copies

In this section, we study the power of quantum secret sharing when we have access to multiple copies of the secret. We define quantum secret sharing with multiple copies analogously to the definitions in Section 2.6, with a minor change: we require that the share function QSS.Share takes as input $|\psi\rangle^{\otimes k}$ where $|\psi\rangle$ is the secret and $k$ is the number of copies we have access to. We give an upper bound, as a function of $t$, on the number of copies required to obtain *efficient* computational quantum secret sharing schemes for *all* $t$-heavy monotone functions (equivalently, all monotone functions with minimal authorized sets of size at least $t$) by showing explicit such schemes. This generalizes a result of Chien [Chi20], who non-explicitly showed existence of schemes only for threshold functions and did not consider efficiency and did not give a proof of security. In particular, our result implies that, given $n$ copies of the secret, we can obtain *efficient* computational quantum secret sharing schemes realizing any function in monotone P.

We first start by showing that, given multiple copies, we can perfectly quantum secret share for $\mathsf{Th}_n^t$. While existence of a scheme was already shown by [Chi20], we further show an explicit efficient scheme for any $t$. Then, by observing that any monotone function $f$ is $t$-heavy when we set $t$ to the size of minimum authorized set of $f$, we can use our compiler to get a polynomial size scheme for $f$ as long as there exists a polynomial size classical scheme for it, which is true for any $f \in$ monotone P by Lemma 4.

**Lemma 8.** *For any monotone function $g : \{0,1\}^{n-1} \to \{0,1\}$ and a perfect classical secret sharing scheme of size $c$ realizing it, there is a perfect quantum secret sharing realizing the monotone function $h(x) = x_n \cdot g(x)$ using 1 qubit and $2c$ classical bits.*

*Proof.* We can use our compiler (Theorem 9) with $f(x) = x_n \cdot g(x_1, \ldots, x_{n-1})$ and $f'(x) = x_n$, along with the naive code QC realizing $f'$ where we give the state directly to the $n$-th party. □

**Lemma 9.** *For any $t \geq 1$, there is an efficient perfect quantum secret sharing of size $O(n^2 \log(n))$ realizing $\mathsf{Th}_n^t$ using $\max(n - 2t + 2, 1)$ copies of the secret.*

*Proof.* The case where $t > \frac{n}{2}$ follows from Theorem 2, so assume $t \leq \frac{n}{2}$. Observe that

$$\mathsf{Th}_n^t = \mathsf{Th}_{n-1}^t + x_n \cdot \mathsf{Th}_{n-1}^{t-1}.$$

We will first show a valid quantum erasure correcting code $\mathsf{QC}$ for $\mathsf{Th}_n^t$. Then, we will use our compiler with $\mathsf{QC}$ and classical Shamir's scheme to get a perfect QSS scheme for $\mathsf{Th}_n^t$.

Suppose we can implement a QECC for $\mathsf{Th}_n^t$ using $c(n,t)$ copies of the secret, with share size $q(n,t)$ qubits and $b(n,t)$ classical bits. First, observe that by invoking Lemma 8 with the classical Shamir's scheme, we get a scheme for $x_n \cdot \mathsf{Th}_{n-1}^{t-1}$ that takes as input a single copy and has classical share size $O(n \log n)$ and qubit share size 1. Therefore, by separately running the scheme for $\mathsf{Th}_{n-1}^t$ using $c(n-1, t)$ copies and running the scheme for $x_n \cdot \mathsf{Th}_{n-1}^{t-1}$ using a single copy, we get a valid QECC for $\mathsf{Th}_n^t$. Hence,

$$c(n, t) = c(n - 1, t) + 1,$$
$$q(n, t) = q(n - 1, t) + 1,$$
$$b(n, t) = b(n - 1, t) + O(n \log n).$$

As the base case, by Lemma 5, we have

$$c(2t - 1, t) = 1,$$
$$q(2t - 1, t) = O(n \log n),$$
$$b(2t - 1, t) = 0.$$

Therefore, we recursively apply the construction above to get a QECC for $\mathsf{Th}_n^t$. Finally, we use our compiler with this resulting QECC and the classical Shamir's scheme to get a perfect QSS scheme for $\mathsf{Th}_n^t$ that uses $n - 2t + 2$ copies of the secret and has qubit share size $O(n \log n)$ and classical share size $O(n^2 \log n)$. □

**Theorem 14** (Theorem 5, restated). *A total of $\max(1, n - 2t + 2)$ copies of the quantum secret are sufficient to obtain efficient computational quantum secret sharing schemes realizing all $t$-heavy monotone functions $f$ in monotone $\mathsf{P}$ assuming the existence of post-quantum secure one-way functions.*

*Proof.* Observe that if $f$ is $t$-heavy, then $\mathsf{Th}_n^t \geq f$. Since $f \in$ monotone $\mathsf{P}$, we can invoke the efficient classical scheme from Lemma 4. Plugging this scheme as the classical scheme and the quantum secret sharing scheme from Lemma 9 as the code in our compiler in Theorem 9 yields an efficient computational quantum secret sharing scheme realizing $f$ using the desired number of copies. □

**Corollary 3** (Corollary 2, restated). *For any monotone function $f : \{0, 1\}^n \to \{0, 1\}$ in monotone $\mathsf{P}$ there is an efficient computational quantum secret sharing scheme using at most $n$ copies of the secret realizing $f$ based on the existence of post-quantum secure one-way functions.*

*Proof.* Observe that any $f$ is $t$-heavy for $t = \min_{P \subseteq [n] : f(P) = 1} |P| \geq 1$. Invoke Theorem 5 to obtain the result. □

The proof of Theorem 6 is analogous to the proof of Theorem 5 above, except that we use the classical secret sharing scheme from Lemma 7 instead.

**Remark 1.** *It is known that quantum secret sharing schemes with multiple copies which feature independent sharing and reconstruction procedures for each copy of the secret require at least* $\max(n - 2t + 2, 1)$ *copies of the secret to realize the threshold function* $\mathsf{Th}_n^t$ *[Chi20], which is t-heavy. Therefore, improving our result above would require a fundamentally different approach.*

## 9 Heavy monotone functions are as hard as arbitrary monotone functions

We have shown that we can lift many important classical results to the quantum setting for all heavy monotone functions. In this section, we argue formally that heavy monotone functions (recall Definition 1) inherit important hardness properties of arbitrary monotone functions, as captured in Proposition 1. In particular, some heavy functions in monotone P require exponential size monotone span programs.

**Proposition 4** (Proposition 1, restated)**.** *Let* $\mathsf{mSP}(f)$ *and* $\mathsf{mC}(f)$ *denote the size of the smallest monotone span program and monotone circuit computing $f$, respectively. Then, for every monotone function* $f : \{0,1\}^n \to \{0,1\}$ *there exists a heavy monotone function* $f' : \{0,1\}^{2n} \to \{0,1\}$ *such that* $\mathsf{mSP}(f') \geq \frac{\mathsf{mSP}(f)}{2n}$ *and* $\mathsf{mC}(f') \leq \mathsf{mC}(f) + n$. *Moreover, if $f$ is in* mNP*, then so is $f'$.*

*Proof.* Fix a monotone function $f : \{0,1\}^n \to \{0,1\}$ and consider the associated monotone function $f' : \{0,1\}^{2n} \to \{0,1\}$ defined by

$$f'(x_1, \ldots, x_{2n}) = f(x_1, \ldots, x_n) \cdot x_{n+1} \cdots x_{2n}.$$

It is clear that $f'$ is heavy, since $f'(P) = 1$ implies that $\{n+1, \ldots, 2n\} \subseteq P$ and that $P$ contains at least one of $\{1, \ldots, n\}$. Moreover, it is also easy to see that if $f$ is in mNP, then so is $f'$.

We now proceed to relate the sizes of the smallest monotone span programs and circuits computing $f$ and $f'$. Let $M'$ be an MSP computing $f'$. For $i \in [n]$, define $M'_i$ as the matrix consisting of rows of $M'$ labeled by $x_i$, and let $M'_{n+1}$ be the matrix consisting of rows labeled by $x_{n+1}, \ldots, x_{2n}$. Consider the MSP $M$ obtained by vertically stacking $M'_1, M'_{n+1}, M'_2, M'_{n+1}, \ldots M'_n, M'_{n+1}$, and labeling their rows $1, 1, 2, 2, \ldots, n, n$ respectively. Then, it is easy to see that $M$ computes $f$. Observing that $\mathsf{MSP.size}(M') \leq 2n \cdot \mathsf{MSP.size}(M)$ yields the desired bound

$$\mathsf{mSP}(f) \leq 2n \cdot \mathsf{mSP}(f').$$

To see the bound for monotone circuits, note that we can test whether $x_{n+1} = \cdots = x_{2n} = 1$ via a tree composed of less than $n$ AND gates. Then, we can just compute an AND between the output of this tree and the output of $f$. Therefore, it follows that

$$\mathsf{mC.size}(f') \leq \mathsf{mC.size}(f) + n. \qquad \square$$

We combine this proposition with recent strong separation results between monotone circuits and span programs [RPRC16, PR17].

**Lemma 10** ([RPRC16])**.** *There exist functions in* monotone P *which require monotone span programs of size* $\exp(n^{\Omega(1)})$.

Combining Proposition 1 and Lemma 10 immediately yields the following corollary.

**Corollary 4** (Corollary 1, restated)**.** *There exist heavy monotone functions in* monotone P *which require monotone span programs of size* $\exp(n^{\Omega(1)})$.

Remarkably, this shows that there exist heavy monotone functions that can be realized by efficient computational *classical* secret sharing schemes based on standard hardness assumptions, but for which current methods for constructing quantum secret sharing schemes can only yield schemes with exponential share size.

## 10 Beyond heavy monotone functions

In this section, we discuss how to use our compiler to lift classical schemes for two natural extensions of heavy functions, which are provably richer families than the class of heavy functions.

### 10.1 Weighted heavy monotone functions

We start with weighted heavy functions, a natural extension of heavy functions analogous to weighted threshold functions.

**Definition 16** (Definition 2, restated). *Let $w : [n] \to \mathbb{N}$ be an integer weight function. A monotone function $f : \{0,1\}^n \to \{0,1\}$ is said to be $(w,t)$-weighted-heavy if for any set $P \subseteq [n]$ with $f(P) = 1$ we have $\sum_{i \in P} w(i) \geq t$. Moreover, we call $W = \sum_{i=1}^n w(i)$ the total weight of $f$. If $f$ is $(w,t)$-weighted-heavy for some $w$ and $t = \lfloor \frac{W}{2} \rfloor + 1$, we simply call it $w$-weighted-heavy or just weighted-heavy if $w$ is clear from context.*

Observe that when $t > \frac{W}{2}$, a $(w,t)$-weighted heavy function is necessarily no-cloning and hence admits a quantum secret sharing scheme. We begin by showing that the family of weighted-heavy monotone functions with polynomial total weight strictly generalizes the classes of weighted-heavy monotone functions and weighted threshold functions.

**Proposition 5** (Proposition 2, restated). *There are families of $w$-weighted heavy monotone functions with total weight $W = \mathrm{poly}(n)$ which are neither heavy nor weighted threshold functions with $\mathrm{poly}(n)$ weights. Moreover, there exist such functions which are also in* monotone P *but require monotone span programs of size $\exp(n^{\Omega(1)})$.*

*Proof.* Consider the family of monotone functions $f : \{0,1\}^n \to \{0,1\}$ given by

$$f(x_1, \ldots, x_n) = \mathsf{Th}_3^2(x_1, x_2, g(x_3, \ldots, x_n)),$$

where $g : \{0,1\}^{n-2} \to \{0,1\}$ is an arbitrary heavy monotone function and $\mathsf{Th}_3^2$ denotes the 2-out-of-3 threshold function. It is clear that no such $f$ is heavy since $\{1,2\}$ is authorized. Also observe that $f$ is $w_f$-heavy for

$$w_f(i) = \begin{cases} n-3, & \text{if } i = 1 \text{ or } i = 2, \\ 1, & \text{if } i \geq 3, \end{cases}$$

which satisfies $W = 3n - 9$. We claim that $f$ cannot be written as a weighted threshold function with $\mathrm{poly}(n)$ weights whenever the same holds for $g$.

To see this, suppose for a contradiction that $f$ is a weighted threshold function with weight function $w$ and threshold $t$. For $P \subseteq \{3, \ldots, n\}$, let $w(P) = \sum_{i \in P} w(i)$. Then, if $g(P) = 1$ it must be the case that

$$w(1) + w(P) \geq t \quad \text{and} \quad w(2) + w(P) \geq t,$$

and so $w(P) \geq t - \min(w(1), w(2)) \geq t - \frac{w(1)+w(2)}{2}$. On the other hand, if $g(P) = 0$ then

$$w(1) + w(P) < t \quad \text{and} \quad w(2) + w(P) < t,$$

and so $w(P) < t - \max(w(1), w(2)) \leq t - \frac{w(1)+w(2)}{2}$. Therefore, $g$ would have to be a weighted threshold function with weight function $w$ and threshold $t' = t - \frac{w(1)+w(2)}{2}$.

Finally, pick $g$ to be a heavy monotone function in monotone P that requires monotone span programs of size $\exp\left(n^{\Omega(1)}\right)$, whose existence is guaranteed by Corollary 1. Then, it is easy to see that $f$ is also in monotone P. To see that $f$ also requires monotone span programs of size $\exp\left(n^{\Omega(1)}\right)$, observe the following. Consider any input such that $x_1 = 0, x_2 = 1$. Then, we get $f(x_1, \ldots, x_n) = g(x_3, \ldots, x_n)$. Therefore, since threshold gates can be computed by polynomial size monotone span programs, if there was a polynomial size MSP computing $f$, we could obtain a polynomial size MSP computing $g$ by this hardwiring construction, which is a contradiction. Finally, observe that weighted threshold functions with poly($n$) weights have polynomial size MSPs, and therefore $g$ cannot be such a function. Then, by above, $f$ cannot be a weighted threshold function with polynomial weights, as desired. $\square$

Observe that we have proved that there are weighted-heavy monotone functions with polynomial total weight, outside the classes of heavy monotone functions and weighted threshold functions with polynomial total weight which are in monotone P but require exponentially large monotone span programs. Hence, neither previously known perfect quantum schemes [Got00, Smi00] nor our results above on computational quantum secret sharing for heavy monotone functions (such as Theorem 2) yield efficient schemes for such functions. Nevertheless, below we use our compiler to construct efficient quantum secret sharing schemes for weighted-heavy monotone functions.

**Theorem 15** (Theorem 7, restated). *If $f : \{0,1\}^n \to \{0,1\}$ is a $(w,t)$-weighted-heavy monotone function in monotone P with total weight $W = \text{poly}(n)$ and threshold $t > W/2$, then there is an efficient computational quantum secret sharing scheme realizing $f$ based on the existence of post-quantum secure one-way functions.*

*Proof.* Let $f'$ be the $\left(w, t = \left\lfloor \frac{W}{2} \right\rfloor + 1\right)$-weighted-threshold function such that $f'(P) = 1$ if and only if $\sum_{i \in P} w(i) \geq t$. Note that $f' \geq f$. We construct an efficient QC realizing $f'$ as follows. Consider the quantum Shamir secret sharing scheme from Lemma 5 realizing $\text{Th}_W^t$. Then, give any $w(i)$ shares of this scheme to the $i$-th party. It is clear that this QC realizes $f'$. Furthermore, it is efficient since $W = \text{poly}(n)$. We let SS be Yao's scheme from Lemma 4 realizing $f$, which is efficient since $f \in$ monotone P. Invoking Theorem 9 for $f$ with this choice of $f'$, QC, and SS gives us an efficient quantum secret sharing scheme realizing $f$. $\square$

## 10.2 Trees of weighted-heavy monotone functions

We now show that our results can be generalized even further. In this section, we consider the class of monotone functions $f : \{0,1\}^n \to \{0,1\}$ that can be represented as a constant-depth tree consisting of gates computing weighted-heavy functions of fan-in at most $n$ and poly($n$) total weight. We first show that this class of monotone functions strictly generalizes weighted-heavy monotone functions. This is true already for depth-2 trees.

**Proposition 6** (Proposition 3, restated). *There are families of functions $f : \{0,1\}^n \to \{0,1\}$ which satisfy all of the following at once:*

- *$f$ is not weighted-heavy;*
- *$f \in$ monotone P;*

- *f requires monotone span programs of size $\exp(n^{\Omega(1)})$;*
- *f can be represented as a depth-2 polynomial size tree of weighted-heavy monotone functions where each gate is in* monotone P *and has polynomially bounded total weight.*

*Proof.* Assume that $n$ is a multiple of 9 and define $k = \frac{n}{3}$. Let $g : \{0,1\}^n \to \{0,1\}$ be a heavy monotone function in monotone P that requires monotone span programs of size $\exp(n^{\Omega(1)})$. Note that the existence of such $g$ is guaranteed by Corollary 1. For ease of the calculations we will show a function $f$ on $2n$ bits, but this is without loss of generality. Consider the function

$$f(x_1, \ldots, x_{2n})$$
$$= \mathsf{Th}_{k+1}^{\frac{2k}{3}}(\mathsf{Th}_3^2(x_1, x_2, x_3), \ldots, \mathsf{Th}_3^2(x_{3i-2}, x_{3i-1}, x_{3i}), \ldots, \mathsf{Th}_3^2(x_{n-2}, x_{n-1}, x_n),$$
$$g(x_{n+1}, \ldots, x_{2n})),$$

which is computed by a depth-2 tree of heavy functions, where each gate has total weight bounded polynomially and also each gate is in monotone P.

We claim $f$ is not weighted-heavy. Consider any weight function $w : [2n] \to \mathbb{N}$ for $f$. Define the permutation $\ell : [k] \to [k]$ so that the sum of weights for the $j$-th sub-gate,

$$w(3j-2) + w(3j-1) + w(3j),$$

is the $\ell(j)$-th smallest among all $k$ such sums for all $j \in [k]$. Furthermore, for each $j \in [k]$, define $q_j = \operatorname{argmax}_{i \in \{3j-2, 3j-1, 3j\}} w(i)$. Then, consider the input $x \in \{0,1\}^{2n}$ defined as

$$x_i = \begin{cases} 0, & \text{if } i > n \text{ or } \ell(\lfloor \frac{i+2}{3} \rfloor) > \frac{2k}{3} \text{ or } q_{\lfloor \frac{i+2}{3} \rfloor} = i, \\ 1, & \text{otherwise}, \end{cases}$$

for all $i \in [2n]$. It is easy to see that $f(x) = 1$. Observe that $x$ is defined so that (i) heaviest $\frac{k}{3}$ threshold sub-gates get all 0 as their input, (ii) for the remaining $\frac{2k}{3}$ threshold sub-gates, the heaviest input is 0 and the other two inputs are 1 and (iii) $g$ gets all 0 as its input. As a result, we have that

$$\sum_{i=1}^{2n} x_i w(i) = \sum_{j=1}^{\frac{2k}{3}} \sum_{\delta=-2}^{0} x_{\ell^{-1}(j+\delta)} w(\ell^{-1}(j+\delta))$$
$$\leq \frac{2}{3} \sum_{j=1}^{\frac{2k}{3}} \sum_{\delta=-2}^{0} w(\ell^{-1}(j+\delta))$$
$$\leq \frac{2}{3} \cdot \frac{2}{3} \sum_{i=1}^{n} w(i)$$
$$\leq \frac{4}{9} \sum_{i=1}^{2n} w(i)$$
$$< \frac{1}{2} \sum_{i=1}^{2n} w(i).$$

Therefore, $f$ is not $w$-weighted-heavy.

It is easy to see that $f \in$ monotone P whenever $g \in$ monotone P. To see $f$ requires monotone span programs of size $\exp\left(n^{\Omega(1)}\right)$, observe the following. Consider any input $x \in \{0,1\}^{2n}$ that satisfies $x_i = 1$ for all $i \leq 2k-3$ and $x_i = 0$ for all $2k-2 \leq i \leq n$. Then, we get $f(x_1, \ldots, x_{2n}) = g(x_{n+1}, \ldots, x_{2n})$. Therefore, since threshold gates can be computed by polynomial size monotone span programs, if there was a polynomial size MSP computing $f$, we could obtain a polynomial size MSP computing $g$ by this hardwiring construction, which is a contradiction.

□

This proposition in particular shows that neither our previous results (such as Theorem 2 and Theorem 7) nor the previously known perfect quantum secret sharing schemes from [Got00, Smi00] directly yield efficient schemes for constant-depth trees of weighted heavy monotone functions with polynomial total weight. Yet, we can once again exploit our previous results to obtain efficient schemes for this more general class of functions.

**Theorem 16** (Theorem 8, restated)**.** *Let $\mathcal{F}$ be the family of weighted heavy functions in* monotone P *with* poly($n$) *total weight and fan-in at most $n$. If $f : \{0,1\}^n \to \{0,1\}$ is computed by a constant depth polynomial size tree composed of gates in $\mathcal{F}$, then there is an efficient computational quantum secret sharing scheme realizing $f$ based on the existence of post-quantum secure one-way functions.*

*Proof.* We will follow a gate by gate approach, analogous to [Yao89, VNS$^+$03]. Let $T$ be the tree computing $f$, as described in the theorem statement. We start by placing the single qubit secret state $\rho$ on the output wire. Then, we share $\rho$ according to the top gate using the efficient quantum secret sharing scheme from Theorem 7, giving the shares to each input wire coming to this gate. Then, we recursively repeat this for each gate below, re-sharing all the shares that gate has obtained using Theorem 7, until we reach wires that correspond to inputs $x_i$ of $f$, whereby we give the corresponding shares to party $i$.

Note that this recursive procedure is efficient because we secret share for at most polynomially many gates and for each gate in this process (i) secret sharing for that gate is efficient by Theorem 7, (ii) the secret to be shared is of size poly($n$), since the tree has constant depth and the gates have fan-in at most $n$. Observe also that this constitutes a valid QECC QC realizing $f$ of polynomial size. Then, we let SS be the efficient computational classical secret sharing scheme from Lemma 4 realizing $f$ based on the existence of post-quantum secure one-way functions. Finally we invoke our compiler (Theorem 9) for $f$ with $f' = f$, the QECC QC above, and SS to obtain an efficient computational quantum secret sharing scheme realizing $f$ based on the existence of post-quantum secure one-way functions.

□

## Acknowledgements

Research supported by the following grants of Vipul Goyal: NSF award 1916939, DARPA SIEVE program, a gift from Ripple, a DoE NETL award, a JP Morgan Faculty Fellowship, a PNC center for financial services innovation award, and a Cylab seed funding award. João Ribeiro's research was also supported by NOVA LINCS (UIDB/04516/2020) with the financial support of FCT - Fundação para a Ciência e a Tecnologia.

# Appendix

## A  Quantum information theory

In this section, we present useful properties of the distance measures for quantum information introduced in Section 2.2.

**Lemma 11** (Trace distance). *For any two density matrices $\rho, \sigma$ the following holds:*

i. *For any trace-preserving quantum operation $\Phi$,*
$$D(\Phi(\rho), \Phi(\sigma)) \leq D(\rho, \sigma)$$

ii. *Assuming the states are of a composite system $AB$,*
$$D(\rho^A, \sigma^A) \leq D(\rho^{AB}, \sigma^{AB})$$

iii. *For any two probability distributions $\{p_i\}_{i \in I}, \{q_i\}_{i \in I}$, and ensembles of states $\{\rho_i\}_{i \in I}, \{\sigma_i\}_{i \in I}$,*
$$D\left(\sum_{i \in I} p_i \rho_i, \sum_{i \in I} q_i \sigma_i\right) \leq \Delta(p_i, q_i) + \sum_{i \in I} p_i D(\rho_i, \sigma_i)$$

iv. *For a probability distribution $\{p_i\}_{i \in I}$ and an ensemble of states $\{\rho_i\}_{i \in I}$*
$$D\left(\sum_{i \in I} p_i \rho_i, \sigma\right) \leq \sum_{i \in I} p_i D(\rho_i, \sigma_i)$$

v. *For any state $\tau$,*
$$D(\rho \otimes \tau, \sigma \otimes \tau) = D(\rho, \sigma)$$

vi. *For any states $\tau, \upsilon$,*
$$D(\rho \otimes \tau, \sigma \otimes \upsilon) \leq D(\rho, \sigma) + D(\tau, \upsilon)$$

*Proof.* See [NC10, Section 9.2.1] for proofs of (i), (ii) and (iii). Inequality (iv) is a simple corollary of (iii) when $q_i$ is set to $p_i$ and $\sigma_i$ to $\sigma$. Result (v) can be obtained from (i) by setting $\Phi(\gamma) = \gamma \otimes \tau$. Inequality (vi) can be obtained by applying triangle inequality in combination with (v) to $D(\rho \otimes \tau, \sigma \otimes \tau)$ and $D(\sigma \otimes \tau, \sigma \otimes \upsilon)$. □

The advantage satisfies similar properties as the trace distance.

**Lemma 12** (Advantage). *For any circuit family $\mathcal{F}$ and two states $\rho, \sigma$, the following holds:*

i. $A_\mathcal{F}(\rho, \rho) = 0$;

ii. $A_\mathcal{F}(\rho, \sigma) = A_\mathcal{F}(\sigma, \rho)$;

iii. *For any state $\tau$,*
$$A_\mathcal{F}(\rho, \sigma) \leq A_\mathcal{F}(\rho, \tau) + A_\mathcal{F}(\tau, \sigma);$$

iv. Assuming the states are of a composite system $AB$,
$$A_{\mathcal{F}}(\rho^A \otimes |0\rangle\langle 0|, \sigma^A \otimes |0\rangle\langle 0|) \leq A_{\mathcal{F}}(\rho^{AB}, \sigma^{AB});$$

v. For any two probability distributions $\{p_i\}_{i \in I}, \{q_i\}_{i \in I}$, and ensembles of states $\{\rho_i\}_{i \in I}, \{\sigma_i\}_{i \in I}$,
$$A_{\mathcal{F}}\left(\sum_{i \in I} p_i \rho_i, \sum_{i \in I} q_i \sigma_i\right) \leq \Delta(p_i, q_i) + \sum_{i \in I} p_i A_{\mathcal{F}}(\rho_i, \sigma_i);$$

vi. For a probability distribution $\{p_i\}_{i \in I}$ and an ensemble of states $\{\rho_i\}_{i \in I}$
$$A_{\mathcal{F}}\left(\sum_{i \in I} p_i \rho_i, \sigma\right) \leq \sum_{i \in I} p_i A_{\mathcal{F}}(\rho_i, \sigma_i);$$

vii. For any family $\mathcal{F}'$ and state $\tau$ such that there is $C' \in \mathcal{F}'$ satisfying $C'(\rho) = C(\rho \otimes \tau)$ and $C'(\sigma) = C(\sigma \otimes \tau)$ for any $C \in \mathcal{F}$,
$$A_{\mathcal{F}}(\rho \otimes \tau, \sigma \otimes \tau) \leq A_{\mathcal{F}'}(\rho, \sigma).$$

*Proof.* It is easy to see that (i) and (ii) holds. Inequality (iii) is a result of triangle equality on absolute value. Observe that these show that $A_{\mathcal{F}}$ is a pseudometric.

For (iv), consider a subfamily $\mathcal{F}'$ where the circuits ignore the subsystem $B$ and use $|0\rangle$ instead. For (v), observe that there is a circuit $C^* \in \mathcal{F}$ such that
$$\left|\Pr\left[C^*(\sum_{i \in I} p_i \rho_i) = 1\right] - \Pr\left[C^*(\sum_{i \in I} q_i \sigma_i) = 1\right]\right| = A_{\mathcal{F}}\left(\sum_{i \in I} p_i \rho_i, \sum_{i \in I} q_i \sigma_i\right).$$

Then,
$$A_{\mathcal{F}}\left(\sum_{i \in I} p_i \rho_i, \sum_{i \in I} q_i \sigma_i\right) = \left|\Pr\left[C^*(\sum_{i \in I} p_i \rho_i) = 1\right] - \Pr\left[C^*(\sum_{i \in I} q_i \sigma_i) = 1\right]\right|$$
$$= \left|\sum_{i \in I} p_i \Pr[C^*(\rho_i) = 1] - \sum_{i \in I} q_i C^* \Pr[(\sigma_i) = 1]\right|$$
$$= \left|\sum_{i \in I} p_i \Pr[C^*(\rho_i) = 1] - \sum_{i \in I} p_i C^* \Pr[(\sigma_i) = 1] + \sum_{i \in I} (p_i - q_i) \Pr[C^*(\sigma_i) = 1]\right|$$
$$\leq \left|\sum_{i \in I} p_i \Pr[C^*(\rho_i) = 1] - \sum_{i \in I} p_i C^* \Pr[(\sigma_i) = 1]\right| + \left|\sum_{i \in I} (p_i - q_i) \Pr[C^*(\sigma_i) = 1]\right|$$
$$\leq \sum_{i \in I} p_i A_{\mathcal{F}}(\rho_i, \sigma_i) + \Delta(p_i, q_i).$$

Result (vi) is straightforward from (v).

For (vii), as above consider the circuit $C^* \in \mathcal{F}$ satisfying
$$A_{\mathcal{F}}(\rho \otimes \tau, \sigma \otimes \tau) = |\Pr[C^*(\rho \otimes \tau) = 1] - \Pr[C^*(\sigma \otimes \tau) = 1]|.$$

Then, since there is $C' \in \mathcal{F}'$ such that $C'(\rho) = C^*(\rho \otimes \tau)$ and $C'(\sigma) = C^*(\sigma \otimes \tau)$, we get
$$A_{\mathcal{F}'}(\rho, \sigma) = \max_{C \in \mathcal{F}'} |\Pr[C(\rho \otimes \tau) = 1] - \Pr[C(\sigma \otimes \tau) = 1]|$$
$$\geq |\Pr[C'(\rho \otimes \tau) = 1] - \Pr[C'(\sigma \otimes \tau) = 1]|$$
$$= A_{\mathcal{F}}(\rho \otimes \tau, \sigma \otimes \tau).$$

□

# B    Coding theory

In this section, we state some basic concepts from coding theory.

**Definition 17** (Linear code [GRS22]). *An $[n, k, d]_q$ code $\mathsf{C}$ is a linear subspace of $\mathbb{F}_q^n$ of dimension $k$ with $\min_{c \in \mathsf{C} \setminus \{0\}} wt(c) \geq d$, where $wt(c) = |\{i \in [n] : c_i \neq 0\}|$ denotes the* Hamming weight *of $c$.*

The following lemma is based on Reed-Solomon codes.

**Lemma 13** (Reed-Solomon codes [GRS22]). *For all integers $k \leq n$ and every prime power $q \geq n$ there exists an $[n, k, d = n - k + 1]_q$ linear code with efficient encoding and decoding procedures.*